\documentclass[journal]{IEEEtran}
\usepackage{latexsym}
\usepackage{amsfonts,amssymb,amsmath}
\usepackage{hyperref}
\def\BibTeX{{\rm B\kern-.05em{\sc i\kern-.025em b}\kern-.08em T\kern-.1667em\lower.7ex\hbox{E}\kern-.125emX}}
\usepackage{algorithm}
\usepackage{algpseudocode}
\usepackage{amsmath}
\usepackage{amsfonts}
\usepackage{graphicx}
\usepackage{epsfig}
\usepackage{cite}
\usepackage{float}
\usepackage{txfonts}
\usepackage{relsize}
\usepackage{color}
\usepackage{color}
\usepackage{CJK}
\usepackage{times}
\usepackage{multicol}
\usepackage{stfloats}
\usepackage{float}
\usepackage{lipsum}
\usepackage{bm}

\begin{document}

\title{Fluid Antenna System Meets Low-Resolution ADCs in Energy-Efficient Cell-Free Massive MIMO

\thanks{This work was supported by the Hong Kong Research Grants Council with Area of Excellence grant AoE/E-601/22-R.}}
\author{Jun~Qian,~\IEEEmembership{Member,~IEEE,}
Ross~Murch,~\IEEEmembership{Fellow,~IEEE,}~and~Khaled~B.~Letaief,~\IEEEmembership{Fellow,~IEEE}

\thanks{The authors are with the Department of Electronic and Computer Engineering, The Hong Kong University of Science and Technology,
Hong Kong (e-mail: eejunqian@ust.hk, eermurch@ust.hk, eekhaled@ust.hk).}}

\maketitle
{\begin{abstract}

Cell-free massive multiple-input multiple-output (MIMO) systems offer immense spectral efficiency (SE). However, the substantial power consumption of high-resolution analog-to-digital converters (ADCs) poses a severe bottleneck for 6G network deployment. This paper proposes a novel fluid antenna system (FAS)-enabled architecture to improve energy efficiency (EE) without sacrificing capacity. Specifically, we integrate FAS into cell-free massive MIMO systems to counteract low-resolution ADCs.  
We establish a comprehensive uplink transmission model and derive analytical expressions for SE and EE. These expressions explicitly capture the quantization error under slow fluid antenna multiple access and quantify the benefits of low-resolution ADCs on EE.
Furthermore, we formulate a joint optimization problem to maximize EE performance. To solve this, we develop an efficient alternating optimization framework. This framework leverages the Dinkelbach algorithm-based fractional programming for power control, alongside novel accelerated projected gradient ascent (APGA) algorithms to optimize both continuous FAS positions and discrete ADC bit allocations. 
Numerical results reveal that low-resolution ADCs aggressively compress signals to save hardware power, which inevitably degrades SE but maintains EE. However, FASs can recover this SE loss thanks to their spatial flexibility and significantly boost EE by improving the received signal prior to destructive quantization. Furthermore, optimized power control can prevent quantization-induced multi-user interference, while efficient bit allocation can reduce exponential hardware power. Ultimately, our proposed FAS-enabled system, coupled with efficient power control and bit allocation, effectively improves the system performance and outperforms
traditional fixed-position antennas. It establishes a highly robust and energy-efficient paradigm for 6G networks.

\end{abstract}

}

\maketitle

\section{Introduction}

\IEEEPARstart{C}{ell}-free massive multiple-input multiple-output (MIMO) technology has gained traction as a promising solution for future sixth-generation (6G) communications. It can inherently address the growing demand for ubiquitous connectivity and higher data rates \cite{7827017,9665300,8097026}.
In essence, cell-free massive MIMO systems consist of numerous geographically distributed access points (APs) connected to a central processing unit (CPU) via backhaul links, jointly and simultaneously serving users located in the area without cell boundaries \cite{7917284,9665300,11196010}. Consequently,
cell-free massive MIMO systems can provide rich macro-diversity gains to significantly enhance spectral efficiency (SE), introduce ubiquitous coverage, and eliminate inter-cell interference\cite{7827017,8097026,11196010}. Concurrently, cell-free massive MIMO systems could surpass conventional small-cell systems by $95\%$-likely SE per user \cite{7827017}, and introduce additional spatial degrees of freedom (DoFs) to improve spatial multiplexing gains and alleviate the impact of non-ideal practical factors\cite{10032129,9875036}. Meanwhile, favorable propagation and channel hardening can characterize the propagation environment of cell-free massive MIMO systems, facilitating simple precoding design and interference management, such as scalable CPU-based large-scale fading decoding (LSFD)\cite{10032129,7917284,8845768}.

The primary assumption in the aforementioned cell-free massive MIMO systems\cite{7827017,9665300,8097026,7917284,11196010} involves the deployment of ideal/high-quality hardware at all APs. 
However, employing a large number of APs with ideal/high-quality hardware is overly optimistic, leading to overwhelming hardware costs, increased power consumption, and unsatisfactory
energy efficiency (EE) as network densification progresses
\cite{10858168,10878991,8756265}. Specifically, the power consumed by analog-to-digital converters (ADCs) deployed at the radio frequency (RF) chains of the APs grows exponentially with ADC precision \cite{8811486}. To address these limitations, deploying lower-quality hardware components is a practical solution. For example, using low-resolution ADCs while keeping the number of RF chains unchanged facilitates the design of energy- and cost-efficient cell-free massive MIMO systems \cite{10878991,10858168,8756265,8811486,9799777,9318477,10041946}. Recent studies on cell-free massive MIMO systems with low-resolution ADCs are described in \cite{10878991,10858168,8756265,8811486,9799777,9318477,10902611}. In these works, quantization error is typically modelled using the additive quantization noise model (AQNM). Specifically, \cite{8756265} evaluated the downlink performance of cell-free massive MIMO, where both APs and users were equipped with low-resolution ADCs. \cite{9799777} considered low-resolution ADCs/DACs at APs with downlink millimetre-wave cell-free massive MIMO systems and low-capacity backhaul links. The reduction in ADC resolution has been shown to establish a trade-off between system performance and power consumption \cite{8811486,9799777}. 
\cite{9318477} considered low-resolution ADCs at APs in Rician fading channels, focusing on the weighted max-min power optimization problem. The impact of low-resolution ADCs on scalable cell-free massive MIMO systems was assessed in \cite{10041946}. 
In \cite{10902611}, the authors analyzed the performance and power allocation of 
two-way relay cell-free massive MIMO systems with low-resolution ADCs. \cite{10878991} explored the EE maximization problem for reconfigurable intelligent surface (RIS)-assisted cell-free massive MIMO systems employing low-resolution ADCs.  
Furthermore, \cite{10858168} implemented rate-splitting multiple access to enhance uplink performance in cell-free massive MIMO systems with low-resolution ADCs, addressing both SE and EE maximization. Nevertheless, the resulting
issues, such as SE degradation, efficient beamforming, and power allocation design, still warrant further attention. Recently, fluid antenna systems (FASs) have rapidly developed and gained attention. They can offer unique spatial reconfigurability and additional spatial DoFs \cite{9650760,10318134}. Therefore, investigating whether FASs can compensate for the performance loss caused by low-resolution ADCs, while maintaining energy-efficient transmissions, is of great interest.

FASs, also known as movable antennas, can harness spatial resources to enhance spatial diversity and multiplexing gains as a forward-looking approach for 6G wireless networks \cite{9650760,10318134,10146262}. This technology features a software-controllable fluidic, conductive or dielectric structure. It contains a single RF chain and multiple preset positions (designated as ports) within a specific region, enabling the optimization of antenna positions to reconfigure channels and allow for extensive connectivity \cite{10328751,10318134,9650760}. Constructing FASs with digitally controlled pixel antennas has also been considered a practical FAS design \cite {10740058}. Moreover,
\cite{9650760} presented fluid antenna multiple access for multi-user communications, achieving notable reliability and rates. In a more practical context, the slow fluid antenna multiple access (s-FAMA) was proposed in \cite{10066316}, updating FAS positions only when the fading channels change, thus advancing beyond impractical symbol-by-symbol updating. The authors of \cite{10103838} incorporated sufficient parameters for precisely approximating the channel distribution of FAS systems to refine the channel model.
Motivated by the flexibility and practicality of FASs, recent work has increasingly focused on integrating FASs into diverse and promising applications \cite{10318134,10146262,10707252}. Specifically, 
 \cite{10318134} analyzed the integration of FASs in both orthogonal multiple access and non-orthogonal multiple access networks, featuring optimal position selection and power allocation. Concurrently, \cite{10146262} introduced a FAS-enabled multi-user interference mitigation strategy in the RIS-assisted rich scattering environment. \cite{10707252} advanced the use of FASs into integrated
sensing and communication systems by optimizing beamforming and antenna positions.

Given the flexibility and practicality of FASs, the identified advantages have sparked considerable interest in enhancing cell-free massive MIMO systems\cite{10694457,10967080,10891142,10827177,11018493}. Currently, \cite{10694457} applied FASs at APs, and \cite{10967080} applied them to users to investigate how their use could benefit system performance. \cite{10891142} introduced a six-dimensional movable
antenna-aided framework to leverage micro and macro spatial diversities in cell-free massive MIMO systems. \cite{10827177} studied the secure transmission of cell-free symbiotic radio systems with movable antenna-equipped APs. \cite{11018493} introduced a novel optimization algorithm to jointly optimize transmit beamforming and movable antenna positions in movable
antenna-aided cell-free massive MIMO systems.
Despite these advancements, the investigation of FASs in cell-free massive MIMO systems remains in its infancy, with current work predominantly focusing on high-resolution ADCs at the APs. Although low-resolution ADCs could improve EE, the inevitable quantization error might
substantially deteriorate the received signal-to-interference-plus-noise (SINR) to cause SE reduction. This raises concerns about whether adopting FASs can
maintain their advantages in terms of SE and EE.
Motivated by these observations, we aim to bridge the understanding gap by analyzing the uplink performance of FAS-enabled cell-free massive MIMO systems with low-resolution ADCs and investigating the role of FASs in enhancing system EE while mitigating SE reductions caused by low-bit quantization errors.

In this paper, we analyze the uplink SE and EE performance of cell-free massive MIMO systems, where users are equipped with FASs and APs utilize low-resolution ADCs. We first establish the uplink data transmission model with AQNM and apply local minimum mean square error (MMSE) precoding to evaluate the sum SE. Then, we introduce the uplink power consumption model to quantify the benefits of low-resolution ADCs on EE. 
Additionally, we address an alternating optimization (AO)-based joint optimization problem to optimize EE, involving power control,
FAS position selection and bit allocation. Then, the key contributions of this work are as follows.
\begin{itemize}
    \item We propose an uplink channel model for the FAS-enabled cell-free massive MIMO system. This model incorporates low-resolution ADCs with quantization errors at the APs. Unlike traditional fixed-position antennas, this approach introduces dynamic spatial flexibility. It leverages the additional spatial DoFs provided by adjustable FAS positions.

\item We derive analytical expressions for SE and EE to investigate the impact of low-resolution ADCs. These expressions incorporate the uplink power consumption model and assume local MMSE precoding. Furthermore, we systematically quantify the EE variations caused by quantization errors and evaluate the comparative spatial gains unlocked by FASs.

  \item We formulate a joint optimization problem encompassing power control, FAS position selection, and bit allocation to maximize EE. Specifically, we utilize a Dinkelbach algorithm-based fractional programming (FP) algorithm for power control, alongside novel accelerated projected gradient ascent (APGA) algorithms\cite{NIPS2015_f7664060,9217298} for FAS position and bit allocation optimization.

    \item Numerical results validate the proposed system performance and demonstrate the efficacy of the proposed optimization algorithms. While low-resolution ADCs degrade sum SE to improve EE, the designed FAS position optimization can compensate for the SE degradation caused by quantization errors and introduce higher EE. Additionally, efficient power control design and bit allocation contribute significantly to improving EE performance when FASs are adopted.

\end{itemize}
The remainder of this paper is organized as follows. Section II develops the cell-free massive MIMO channel model with FAS-enabled users and low-resolution ADCs at APs. Section III formulates the uplink data transmission model and provides analytical expressions of the sum SE and EE. Section IV details the EE maximization problem regarding the joint optimization of power control, FAS position and bit allocation. Section V offers numerical results and discussions. Section VI concludes the paper and outlines future research.

\textit{Notation:} In this work, we utilize ${\textbf H}^T$, ${\textbf H}^H$, ${\textbf H}^*$ and ${\textbf H}^{-1}$ to represent the transpose, conjugate-transpose, conjugate and inverse of a matrix $\textbf H$, respectively. Moreover, ${\textbf I}_N$ denotes an $N \times N$ identity matrix. In addition, $|\cdot|$ and $||\cdot||$ denote the respective Absolute value and Standard norm. Meanwhile, $\text{diag}(\textbf{H} )$ refers to a diagonal matrix containing the diagonal of the square matrix $\textbf{H}$ in its main diagonal elements. 

\section{System Model}
\begin{figure}[t!]
    \centering
    \includegraphics[width=1\linewidth]{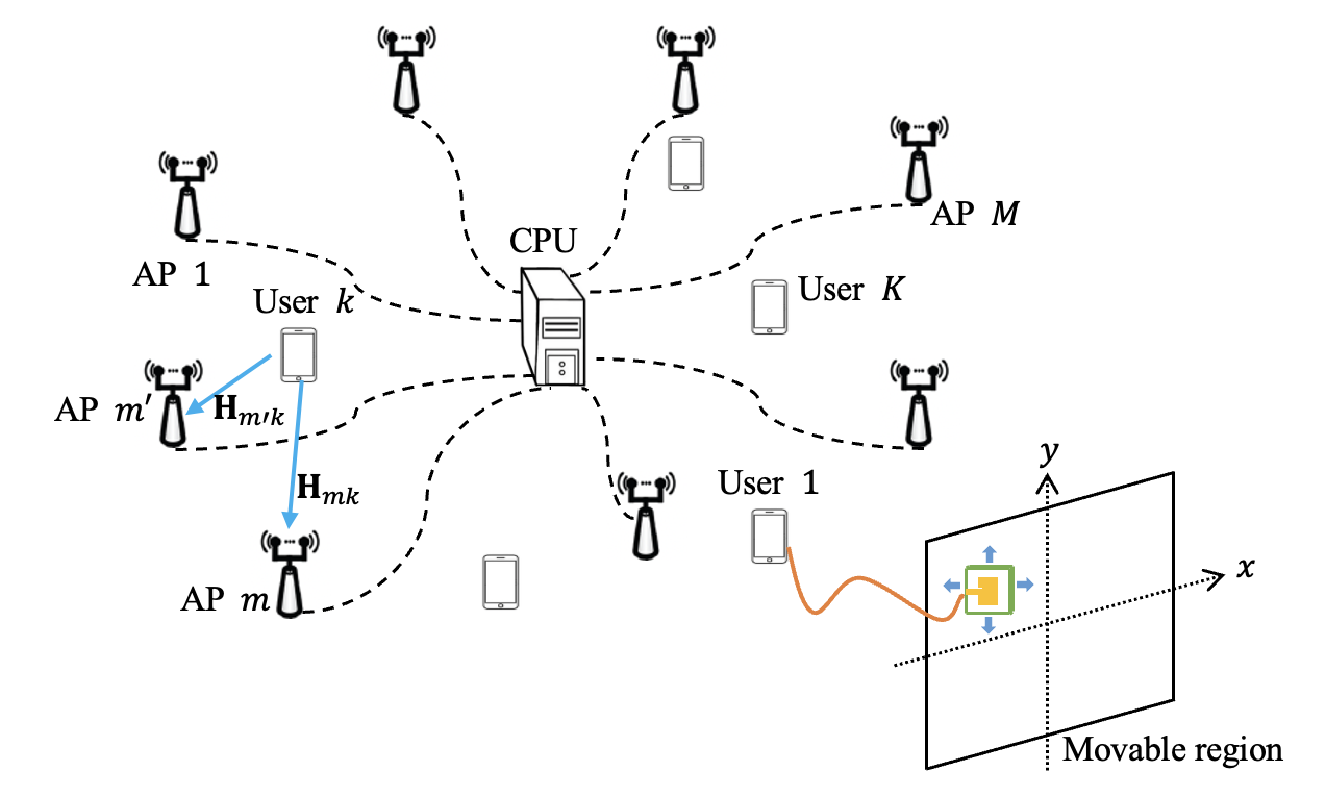}
    \caption{System model of FAS-enabled cell-free massive MIMO.}
    \label{fig_1}
\end{figure}
\subsection{FAS-enabled Channel Model}
As depicted in Fig. \ref{fig_1}, we consider a cell-free massive MIMO system operating in time-division duplex (TDD) mode, comprising $M$ APs and $K$ single-antenna users randomly distributed across a geographic area\cite{10878991,10858168,11196010}, with all APs linked to the CPU via ideal backhaul links. Each AP is equipped with
a uniform planar array possessing $N=N_h \times N_v$ antennas, where
$N_h$ and $N_v$ denote the number of antennas along the horizontal
and vertical directions, respectively\cite{10827177,8388873,10146262}. 
Moreover, each user is equipped with a single two-dimensional (2D) surface FAS \cite{10827177,10318061,10328751,11016053}.
The FAS of each user is connected to the
RF chain via a flexible cable. This allows the antenna to be freely adjusted within a local region $\mathcal{C}_k,~\forall k$
\cite{10318061,10328751,11016053}. A 2D Cartesian coordinate system is leveraged to model the antenna
positions. Specifically, the positions of the $n$-th fixed-position antenna at the $m$-th AP and the FAS at the $k$-th user are respectively denoted as $\textbf{t}_{mn}=[x_{mn},y_{mn}]^T$ and $\textbf{u}_{k}=[x_{k},y_{k}]^T$, $\forall m,~\forall n,~\forall k$ \cite{11018493,10318061}.
Then, 
The position collection of fixed-position antennas at the 
$m$-th AP is denoted
by $\textbf{t}_m=[\textbf{t}_{m1},\textbf{t}_{m2},...,\textbf{t}_{mN}]\in\mathbb{R}^{2\times N},~\forall m$. 
Note that the reference points of the $m$-th AP and the $k$-th user are $D_m$ and $O_k$, respectively.
We assume the FAS adjustment region is significantly smaller than the distance between APs and users. Therefore, we adopt the plane-wave-based far-field model to define the field response \cite{10354003,11018493}. Consequently, the angles of departure (AoDs), angles of arrival (AoAs), and path response amplitudes remain constant across different FAS positions\cite{11016053,10328751}. 
Accordingly, the difference in signal propagation distance for the $l$-th path between the $n$-th fixed-position antenna at the $m$-th AP and the
origin $D_m$ of the local coordinate system at the $m$-th AP is\cite{10328751}
\begin{equation}
\rho_{k,mn}^l(\textbf{t}_{mn})=x_{mn}\text{sin}\theta^r_{mk,l}\text{cos}\phi^r_{mk,l}+y_{mn}\text{cos}\theta^r_{mk,l}.
\end{equation}
Then, the receive field response vector between the $k$-th user and the $n$-th antenna at the $m$-th AP, $\textbf{g}_{mk,n}(\textbf{t}_{mn})\in\mathbb{C}^{L_{mk,r}\times1}$ is
\begin{equation}
   \textbf{g}_{mk,n}(\textbf{t}_{mn})=[e^{j\frac{2\pi}{\lambda}\rho_{k,mn}^1(\textbf{t}_{mn})},e^{j\frac{2\pi}{\lambda}\rho_{k,mn}^2(\textbf{t}_{mn})},...,e^{j\frac{2\pi}{\lambda}\rho_{k,mn}^{L_{mk,r}}(\textbf{t}_{mn})}]^T,
\end{equation}
 and the receive field response vector between the $k$-th user and the $m$-th AP, $\textbf{G}_{mk}(\textbf{t}_{m})=[\textbf{g}_{mk,1}(\textbf{t}_{m1}),...,\textbf{g}_{mk,N}(\textbf{t}_{mN})]\in\mathbb{C}^{L_{mk,r}\times N}$.
Meanwhile, the difference of the signal propagation distance for the $l$-th path between the FAS position $\textbf{u}_{k}$ and the origin $O_k$ of the local coordinate system at the $k$-th user is given by\cite{10243545,10328751}
\begin{equation}
\rho_{mk}^l(\textbf{u}_{k})=x_{k}\text{sin}\theta^t_{mk,l}\text{cos}\phi^t_{mk,l}+y_{k}\text{cos}\theta^t_{mk,l}.
\label{rho_mk}
\end{equation}
Accordingly, the transmit field-response vector for the channel between the $k$-th user and the $m$-th AP, $\textbf{f}_{mk}\left(\textbf{u}_k\right)\in\mathbb{C}^{L_{mk,t}\times 1}$, is expressed as
\begin{equation}
    \textbf{f}_{mk}\left(\textbf{u}_k\right)=[e^{j\frac{2\pi}{\lambda}\rho_{mk}^1(\textbf{u}_{k})},e^{j\frac{2\pi}{\lambda}\rho_{mk}^2(\textbf{u}_{k})},...,e^{j\frac{2\pi}{\lambda}\rho_{mk}^{L_{mk,t}}(\textbf{u}_{k})}]^T.
    \label{f_mk}
\end{equation}
Define the path-response matrix $\boldsymbol{\Sigma}_{mk}\in\mathbb{C}^{L_{mk,r}\times L_{mk,t}}$ as the response between all the transmit and receive channels from $D_m$ to $O_k$, where each entry in the $i$-th row and $j$-th column of $\boldsymbol{\Sigma}_{mk}$ corresponds to the response coefficient between the $j$-th transmit path and the $i$-th receive path between the $k$-th user and the $m$-th AP. Then, the uplink channel vector
from the $k$-th user and the $m$-th AP, $\textbf{h}_{mk}\left(\textbf{u}_k\right)\in\mathbb{C}^{N\times 1}$, is given by
\begin{equation}
\textbf{h}_{mk}\left(\textbf{u}_k\right)=\textbf{G}_{mk}^H\boldsymbol{\Sigma}_{mk}\textbf{f}_{mk}\left(\textbf{u}_k\right).
\end{equation}

\section{Uplink Data Transmission and Performance Analysis}
\subsection{Uplink Data Transmission}
During uplink data transmission, all users coherently transmit signals to all APs \cite {11196010}. 
Thus, the received signal at the $m$-th AP, $\textbf{y}_m \in \mathbb{C}^{N\times 1}$ is obtained as
\begin{equation}
\begin{array}{ll}
     \displaystyle \textbf{y}_m  \displaystyle=\sqrt{p_u}\sum\nolimits_{k=1}^K\textbf{h}_{mk}\left(\textbf{u}_k\right)\sqrt{\eta_k}s_k+\textbf{w}_{m},
\end{array}\label{uplink_received_signal}
   \end{equation} 
where $p_u$ is the maximum user transmit power. $s_k\sim \mathcal{CN}(0,1)$ is the $k$-th user's transmit signal, $\eta_k $ is the power control coefficient with $0<\eta_k \leq 1,~\forall k $. $\textbf{w}_{m}\sim\mathcal{CN}(0,\sigma^2\textbf{I}_{N})$ is the additive white Gaussian noise at the $m$-th AP\cite{11196010,7827017,10858168}. To explore the impact of low-resolution ADCs, we focus on the nonuniform quantizer and adopt the AQNM to quantize the output and characterize the quantization imperfections \cite{9318477,10878991,8811486,10858168}. Thus, we can quantize $\textbf{y}_m$ as
 \begin{equation}
\begin{array}{ll}
     \displaystyle \tilde{\textbf{y}}_m  \displaystyle=\zeta_m\textbf{y}_m+\tilde{\textbf{w}}_m=\zeta_m\sqrt{p_u}\sum\nolimits_{k=1}^K{\textbf{h}}_{mk}\left(\textbf{u}_k\right)\sqrt{\eta_k}s_k+\zeta_m\textbf{w}_{m}+\tilde{\textbf{w}}_m,
\end{array}\label{quantized_uplink_received_signal}
   \end{equation}
where $\zeta_m$ denotes the quantization distortion stemming from the ADCs at the $m$-th AP and is determined by the quantization bits, $b_m$. Note that when $b_m\leq 5$, the value of $\zeta_m$ is given by Table \ref{Table_1}. 
\begin{table}[b!]
\caption{Factor $\zeta_m$ when quantization bits $b_m\leq 5,~\forall m$ \cite{10858168,10878991,8811486}}
\begin{center}
\begin{tabular}{| c | c | c | c |c | c |} 
\hline
\emph {$b_m$} & \emph{1} & \emph{2} & \emph{3} & \emph{4} & \emph{5} \\
\hline
  \textbf {$\zeta_m$} & $0.6366$ & $0.8825$ & $0.9655$ & $0.9905$ & $0.9975$ \\
  \hline
  
\end{tabular}\label{Table_1}
 \end{center}
 \vspace{-12pt}
\end{table}
Then, when $b_m> 5$, $\zeta_m\approx1-\frac{\pi\sqrt{3}}{2}2^{-2b_m}$ \cite{10858168,9318477,10878991}. Besides, 
$\tilde{\textbf{w}}_m\in\mathbb{C}^{N\times 1}$ is the additive Gaussian quantization noise at the $m$-th AP, statistically uncorrelated with $\textbf{y}_m$. According to \cite{9318477,10878991,8811486,9123382}, the covariance matrix of $\tilde{\textbf{w}}_m$ is
 \begin{equation}
\begin{array}{ll}
     \displaystyle \textbf{R}_{\tilde{\textbf{w}}_m}=\zeta_m(1-\zeta_m)\text{diag}\left(p_u\sum\nolimits_{k=1}^K\eta_k{\textbf{h}}_{mk}\left(\textbf{u}_k\right){\textbf{h}}_{mk}\left(\textbf{u}_k\right)^H+\sigma^2\textbf{I}_{N}\right).
\end{array}\label{quantized_noise_covariance}
   \end{equation}

Then, the $m$-th AP utilizes the uplink beamforming $\textbf{v}_{mk}\in\mathbb{C}^{1\times N}$ to detect the signal transmitted by the $k$-th user. Subsequently, all APs pass $\check{s}_{mk}\triangleq \textbf{v}_{mk}\tilde{\textbf{y}}_m,~\forall m,~k, $ to the CPU via backhaul links \cite{9665300,8845768}. The CPU sums up all $\check{s}_{mk}$ from all APs \cite{10891142} to obtain $\hat{s}_{k}=\sum\nolimits_{m=1}^M\check{s}_{mk},~\forall k,$ as \eqref{Quantity_ES} at the top of this page for decoding the information of the $k$-th user,
where $\text{DS}_{\text{k}}$ is the desired signal, $\text{UI}_{\text{kk}'}$ is the inter-user interference, $\text{NS}_{\text{k}}$ is the noise, and $\text{QN}_{\text{k}}$ is the quantization noise.
\begin{figure*}[!t]
\begin{equation}
\begin{array}{ll}
\displaystyle \hat{s}_{k}&   \displaystyle =\sum\nolimits_{m=1}^M\textbf{v}_{mk}\Bigg{(}\zeta_m\sqrt{p_u}\sum\nolimits_{k=1}^K\textbf{h}_{mk}\left(\textbf{u}_k\right)\sqrt{\eta_k}s_k+\zeta_m\textbf{w}_{m}+\tilde{\textbf{w}}_m\Bigg{)}\\&=\underbrace {\sqrt{p_u\eta_k}\sum\nolimits_{m=1}^M \zeta_m \textbf{v}_{mk}\textbf{h}_{mk}\left(\textbf{u}_k\right)s_k}_\text {$\text{DS}_{\text{k}}$}+\displaystyle\sqrt{p_u}\sum\nolimits_{k' \neq k}^K \underbrace {\sum\nolimits_{m=1}^M \zeta_m\sqrt{\eta_{k'}}\textbf{v}_{mk}\textbf{h}_{mk'}\left(\textbf{u}_{k'}\right)s_{k'}}_\text {$\text{UI}_{\text{kk}'}$}+\underbrace {\sum\nolimits_{m=1}^M \zeta_m \textbf{v}_{mk}{\textbf{w}}_{m}}_\text {$\text{NS}_{\text{k}}$}+\underbrace {\sum\nolimits_{m=1}^M \textbf{v}_{mk}\tilde{\textbf{w}}_{m}}_\text {$\text{QN}_{\text{k}}$}.
\end{array}\label{Quantity_ES}
   \end{equation}
\vspace{-5pt}
\hrulefill
\end{figure*} 
\subsection{Spectral Efficiency Analysis}
Then, the uplink achievable SE of the $k$-th user can be formulated as
 \begin{equation}
\begin{array}{ll}
   \displaystyle \text{SE}_k=\text{log}_2\left(1+\text{SINR}_k\right),
   \displaystyle
\end{array}\label{SE_k_uplink}
  \end{equation}
where the uplink SINR of the $k$-th user,
characterizing s-FAMA \cite{10146262,10066316}, can be written as \eqref{SINR_k_uplink} at the top of this page.
\begin{figure*}[!t]
 \begin{equation}
\begin{array}{ll}
   \displaystyle \text{SINR}_k&\displaystyle=\frac{  \displaystyle p_u\eta_k\big{|} \sum\nolimits_{m=1}^M \zeta_m \textbf{v}_{mk}{\textbf{h}}_{mk}\left(\textbf{u}_k\right)\big{|}^2}{\big{|}
  \displaystyle\sqrt{p_u}\sum\nolimits_{k' \neq k}^K \sum\nolimits_{m=1}^M \zeta_m\sqrt{\eta_{k'}}\textbf{v}_{mk}{\textbf{h}}_{mk'}\left(\textbf{u}_{k'}\right)s_{k'}\big{|}^2+\big{|}\displaystyle\sum\nolimits_{m=1}^M \zeta_m \textbf{v}_{mk}{\textbf{w}}_{m}\big{|}^2+\big{|}\displaystyle\sum\nolimits_{m=1}^M \textbf{v}_{mk}\tilde{\textbf{w}}_{m}\big{|}^2}\\
  &\displaystyle=\frac{  \displaystyle p_u\eta_k\big{|} \sum\nolimits_{m=1}^M \zeta_m \textbf{v}_{mk}{\textbf{h}}_{mk}\left(\textbf{u}_k\right)\big{|}^2}{  \displaystyle p_u\sum\nolimits_{k' \neq k}^K\eta_{k'}\big{|}
  \displaystyle \sum\nolimits_{m=1}^M \zeta_m\textbf{v}_{mk}{\textbf{h}}_{mk'}\left(\textbf{u}_{k'}\right)\big{|}^2+\sigma^2 \displaystyle \sum\nolimits_{m=1}^M\big{|}  \zeta_m \textbf{v}_{mk}\big{|}^2+  \displaystyle\sum\nolimits_{m=1}^M \textbf{v}_{mk}\textbf{R}_{\tilde{\textbf{w}}_{m}}
  \textbf{v}_{mk}^H
  }
    ,
\end{array}\label{SINR_k_uplink}
   \end{equation}
\vspace{-10pt}
\hrulefill
\end{figure*} 
Note that the above formulas hold for any uplink beamforming methods. 
In this work, we consider the local MMSE precoding to maximize the instantaneous SINR, where each AP estimates the signal of the $k$-th user using the local instantaneous channel state information (CSI), and the local CSI from all APs is equally combined at the CPU to decode its information \cite{8845768,10891142}. Then, the local MMSE precoding can be given by \eqref{MMSE} at the top of this page\footnote{
In this work, we assume perfect instantaneous CSI is available at the APs to establish theoretical performance bounds and isolate the specific EE-SE trade-offs inherent to FAS-enabled networks. In a practical deployment, channel estimation errors will interact with quantization error to decrease the absolute sum SE and EE gains reported herein. However, the fundamental spatial compensation mechanism of FAS remains valid. The design of practical, low-complexity channel estimation frameworks for this architecture, e.g., leveraging compressed sensing or sparse Bayesian learning\cite{10755170,10236898}, as well as the performance analysis and robust optimization under imperfect CSI, are critical and challenging extensions left for our future work.
}.
\begin{figure*}
\begin{equation}
\begin{array}{ll}
     \displaystyle \textbf{v}_{mk}=\sqrt{p_u\eta_{k}}{\textbf{h}}_{mk}\left(\textbf{u}_k\right)^H\left[\begin{array}{ll}
\zeta_m\displaystyle\sum\nolimits_{k'=1}^Kp_u\eta_{k'}{\textbf{h}}_{mk'}\left(\textbf{u}_{k'}\right){\textbf{h}}_{mk'}\left(\textbf{u}_{k'}\right)^H\displaystyle+
(1-\zeta_m)\text{diag}\left(\displaystyle\sum\nolimits_{k'=1}^Kp_u\eta_{k'}{\textbf{h}}_{mk'}\left(\textbf{u}_{k'}\right){\textbf{h}}_{mk'}\left(\textbf{u}_{k'}\right)^H\right)
+\sigma^2\textbf{I}_{N}\end{array}\right]^{-1}.
\end{array}\label{MMSE}
   \vspace{-4 pt}
   \end{equation}
   \vspace{-10pt}
   \hrulefill
   \end{figure*}

\subsection{Energy Efficiency Analysis }
Notably, implementing low-resolution ADCs degrades sum SE but concurrently reduces power consumption. This trade-off is essential for enhancing EE performance. We evaluate the EE performance of the proposed system to substantiate this observation. A generic power consumption model that accounts for the power consumption of the APs, users, and backhaul links will be adopted \cite{10858168,8811486}. According to \cite{10858168,10878991,11091342}, the total power consumption of the proposed system can be formulated as
\begin{equation}
\begin{array}{ll}
     \displaystyle {P}_\text{tot}&\displaystyle=\sum\nolimits_{k=1}^K\left(\frac{1}{\varepsilon_k}p_u\eta_k+P_{k,tc}\right)\\&\displaystyle+\sum\nolimits_{m=1}^M\underbrace{\left(c_mP_{m,\text{AGC}}+2NP_{m,\text{ADC}}+NP_{m,\text{res}}\right)}_{P_{m,tc}}\\&\displaystyle+\sum\nolimits_{m=1}^M\left(
     P_{m,bc}B\sum\nolimits_{k=1}^K\text{SE}_k+P_{m,0}
     \right)
       ,
\end{array}\label{total_power}
   \end{equation}
where $\varepsilon_k\in(0,1]$ is the power amplifier efficiency, $P_{k,tc}$, $P_{m,tc}$ denote the power required to operate the circuit components of the $k$-th user and the $m$-th AP. $P_{m,\text{AGC}}$, $P_{m,\text{ADC}}$, and $P_{m,\text{res}}$ stand for the power consumption at automatic gain control
units, ADCs, and residual units at the $m$-th AP, respectively\cite{10878991,11091342}. Meanwhile, $c_m$ is a binary parameter with $c_m=\min{\left(b_m-1,1\right)},~b_m\geq1,~\forall m$. $P_{m,bc}$ denotes the traffic-dependent backhaul link power of the $m$-th AP, and $P_{m,0}$ is the fixed power consumption of
each backhaul link. $B$ is the system bandwidth\cite{10858168,8811486,10878991}.  The specific formula of $P_{m,\text{ADC}}$ is given as
\begin{equation}
\begin{array}{ll}
     \displaystyle P_{m,\text{ADC}}=2^{b_m}\cdot\text{FOM}_\textit{W}\cdot f_s
       ,
\end{array}\label{P_mADC}
   \end{equation}
where $\text{FOM}_\textit{W}$ and $f_s$ represent the Walden's figure-of-merit and Nyquist sampling rate, respectively\cite{10858168}. Furthermore, the system EE is defined as the ratio of the sum SE to the total power consumption. We can formulate the EE of the proposed system equipped with low-resolution ADCs as \cite{10858168,8388873}
\begin{equation}
\begin{array}{ll}
     \displaystyle \text{EE}=\frac{\displaystyle B\sum\nolimits_{k=1}^K\text{SE}_k}{\displaystyle P_\text{tot}}.
\end{array}\label{EE}
   \end{equation}

\section{Energy Efficiency Maximization}
In this section, we focus on the joint design of the power control coefficients, FAS positions and bit allocations to optimize the EE for the proposed system, addressed by developing an efficient AO-based algorithm. Then, the proposed EE optimization problem can be
modelled as\cite{10878991,9205899}
 \begin{subequations}
  \begin{align}
  P_0:~ &\mathop {\max }\limits_{\{\eta_{k}\},\{\textbf{u}_k\},\{b_m\}}~\text{EE}=\frac{\displaystyle B\sum\nolimits_{k=1}^K\text{SE}_k}{\displaystyle P_\text{tot}}
   \\
   & \text{subject~to} \\
   & \text{SE}_k\geq S_\text{min},~\forall k,\label{SE_cons}\\
   &
   0\leq \eta_k\leq 1,~\forall k, \label{power_control_cons}\\
   &\textbf{u}_k\in \mathcal{C}_k,~\forall k,\label{FAS_position_cons}\\
   &
   b_m\in\{b_\text{min}, b_\text{max}\},~\forall m.\label{bit_cons}
  \end{align}
  \label{EE_optimization}
 \end{subequations}
Note that the EE can be rewritten as
\begin{equation}
\begin{array}{ll}
     \displaystyle \text{EE}=\frac{B}{\displaystyle

\frac{\displaystyle\sum\limits_{k=1}^K\left(\frac{1}{\varepsilon_k}p_u\eta_k+P_{k,tc}\right)+\sum\limits_{m=1}^M\left(P_{m,tc}+P_{m,0}\right)}{\displaystyle\sum\nolimits_{k=1}^K\text{SE}_k}+\sum\nolimits_{m=1}^M
     P_{m,bc}
     }.
\end{array}\label{EE_reformulate}
   \end{equation}
Without loss of generality, maximizing EE is equivalent to minimizing the first term of the denominator of \eqref{EE_reformulate}. As a
result, the optimization problem \eqref{EE_optimization} is equivalent to maximize
\vspace{-5pt}
\begin{subequations}
  \begin{align}
  P_1:~&\mathop {\max }\limits_{\{\eta_{k}\},\{\textbf{u}_k\},\{b_m\}}~\frac{\displaystyle\sum\nolimits_{k=1}^K\text{SE}_k}{\displaystyle\sum\nolimits_{k=1}^K\left(\frac{1}{\varepsilon_k}p_u\eta_k+P_{k,tc}\right)+\sum\nolimits_{m=1}^M\left(P_{m,tc}+P_{m,0}\right)}
   \\
   & \text{subject~to} ~~\eqref{SE_cons},\eqref{power_control_cons},\eqref{FAS_position_cons},\eqref{bit_cons}.
  \end{align}
  \label{EE_optimization_all}
 \end{subequations}
 We find that the objective function in \eqref{EE_optimization_all} is a ratio of
concave and affine functions, while the constraints in \eqref{EE_optimization_all}
are convex. Thus, introducing a new auxiliary variable $\vartheta>0$, we transform the optimization problem \eqref{EE_optimization_all} into a sequence of parametric subtractive-form problems using the Dinkelbach method to reformulate the single-ratio problem \eqref{EE_optimization_all} as\cite{10878991,8314727,11091342}
\vspace{-5pt}
\begin{subequations}
  \begin{align}
   P_2:~&\mathop {\max }\limits_{\{\eta_{k}\},\{\textbf{u}_k\},\{b_m\}}~\displaystyle\sum\nolimits_{k=1}^K\text{SE}_k-\vartheta\bar{P}_\text{tot}
   \\
   & \text{subject~to} ~~\eqref{SE_cons},\eqref{power_control_cons},\eqref{FAS_position_cons},\eqref{bit_cons}.
  \end{align}
  \label{EE_optimization_Dink}
  where $\bar{P}_\text{tot}=\displaystyle\sum\nolimits_{k=1}^K\left(\frac{1}{\varepsilon_k}p_u\eta_k+P_{k,tc}\right)+\sum\nolimits_{m=1}^M\left(P_{m,tc}+P_{m,0}\right)$.
 \end{subequations}
 \subsection{power control}
For given $\textbf{u}_k,~b_m,~\forall k,~m$, the subproblem of the uplink power control can be re-formulated as
\begin{subequations}
  \begin{align}
    P_3:~&\mathop {\max }\limits_{\boldsymbol{\eta}}~\displaystyle\sum\nolimits_{k=1}^K\text{SE}_k\left(\boldsymbol{\eta}\right)-\vartheta\bar{P}_\text{tot}\left(\boldsymbol{\eta}\right)
   \\
   & \text{subject~to} 
   ~~\eqref{SE_cons},\eqref{power_control_cons},
  \end{align}
  \label{EE_optimization_P3}
 \end{subequations}  where $\boldsymbol{\eta}=[\eta_1,...,\eta_K]^T\in\mathbb{C}^{K\times 1}$. The auxiliary variable is iteratively updated by $\vartheta^{*}=\frac{\displaystyle\sum\nolimits_{k=1}^K\text{SE}_k\left(\boldsymbol{\eta}\right)}{\displaystyle\bar{P}_\text{tot}\left(\boldsymbol{\eta}\right)}$ when $\boldsymbol{\eta}$ is held fixed \cite{8314727,10878991}.
This work utilizes the closed-form FP approach to solve the
power control problem\cite{8314727}. Then, we first introduce the uplink SINR as a ratio based on \eqref{SINR_k_uplink}: $\text{SINR}_{k}={A_{k}(\boldsymbol{\eta})}/{B_{k}(\boldsymbol{\eta})}$ with
\begin{equation}
    \begin{array}{ll}
  A_{k}(\boldsymbol{\eta})=   \displaystyle p_u\eta_k\big{|} \sum\nolimits_{m=1}^M \zeta_m \textbf{v}_{mk}{\textbf{h}}_{mk}\big{|}^2,
    \end{array}\label{A_k}
\end{equation}
\begin{equation}
    \begin{array}{ll}
  B_{k}(\boldsymbol{\eta})\displaystyle&\displaystyle=  p_u\sum\nolimits_{k' \neq k}^K\eta_{k'}\big{|}
  \displaystyle \sum\nolimits_{m=1}^M \zeta_m\textbf{v}_{mk}{\textbf{h}}_{mk'}\big{|}^2\\\displaystyle&+\sigma^2 \displaystyle \sum\nolimits_{m=1}^M\big{|}  \zeta_m \textbf{v}_{mk}\big{|}^2+  \displaystyle\sum\nolimits_{m=1}^M \textbf{v}_{mk}\textbf{R}_{\tilde{\textbf{w}}_{m}}
  \textbf{v}_{mk}^H.
    \end{array}\label{B_k}
\end{equation}
 Then, both $A_{k}(\boldsymbol{\eta})$ and $B_{k}(\boldsymbol{\eta})$ are non-negative real function and positive real function of $\boldsymbol{\eta}$, quadratic in structure. Thus, the uplink power optimization problem can be given by \cite{9519163,10256066}
 \begin{subequations}
  \begin{align}
   P_4:~&\mathop {\max }\limits_{\boldsymbol{\eta}}~\sum\nolimits_{k=1}^K\text{log}_2\left(1+\frac{{A_{k}(\boldsymbol{\eta})}}{{B_{k}(\boldsymbol{\eta})}}\right)-\vartheta\bar{P}_\text{tot}\left(\boldsymbol{\eta}\right)
   \\
   & \text{subject~to} \nonumber \\
   & \frac{{A_{k}(\boldsymbol{\eta})}}{{B_{k}(\boldsymbol{\eta})}}\geq 2^{S_\text{min}}-1,~\forall k,
   \\
   &
   0\leq \eta_k\leq 1,~\forall k.
  \end{align}
  \label{Uplink_optimization_P2}
 \end{subequations}
 Since $A_{k}(\boldsymbol{\eta})$ and $B_{k}(\boldsymbol{\eta})$ satisfy the conditions required for the general FP algorithm \cite{9293031,9519163,10256066}, the objective function in \eqref{Uplink_optimization_P2} involves the
summation of logarithm, and we exploit the Lagrangian dual transform\cite{9293031,10256066,10878991}
 \begin{subequations}
  \begin{align}
   P_5:~&\mathop {\max }\limits_{\boldsymbol{\eta},\boldsymbol{\gamma}}~s_1(\boldsymbol{\eta},\boldsymbol{\gamma},\vartheta)
   \\
   & \text{subject~to} \nonumber \\
   & \frac{{A_{k}(\boldsymbol{\eta})}}{{B_{k}(\boldsymbol{\eta})}}\geq 2^{S_\text{min}}-1,~\forall k,
   \\
   &
   0\leq \eta_k\leq 1,~\forall k,
  \end{align}
  \label{EE_optimization_P4}
 \end{subequations}
 where $\boldsymbol{\gamma}=[\gamma_{1},...,\gamma_{K}]^T\in\mathbb{R}^{K\times 1}$ is the set of auxiliary SINR variables and the new objective function \cite{10878991,9519163,10812717} is
 \begin{equation}
\begin{array}{ll}
   \displaystyle s_1(\boldsymbol{\eta},\boldsymbol{\gamma},\vartheta)=\sum\limits_{k=1}^K\left(\text{log}(1+\gamma_{k})-\gamma_{k}+\frac{\displaystyle (1+\gamma_{k})A_{k}(\boldsymbol{\eta})}{A_{k}(\boldsymbol{\eta})+B_{k}(\boldsymbol{\eta})}\right)-\vartheta\bar{P}_\text{tot}\left(\boldsymbol{\eta}\right),
\end{array}\label{s_1}
  \end{equation}
where $s_1(\boldsymbol{\eta},\boldsymbol{\gamma},\vartheta)$ is a concave differentiable function over $\boldsymbol{\gamma}$ if $\boldsymbol{\eta}$ is fixed\cite{10878991}. Thus, when $\boldsymbol{\eta}$ is held fixed, we can find the local optimum of $\boldsymbol{\gamma}$ by
solving $\partial s_1(\boldsymbol{\eta},\boldsymbol{\gamma},\vartheta)/{\partial \gamma_{k}}=0$ to obtain the optimal $\gamma_{k}^*=\frac{\displaystyle A_{k}(\boldsymbol{\eta})}{\displaystyle B_{k}(\boldsymbol{\eta})}$ \cite{9293031,10256066,10812717}. Then, to deal with the fractional terms involved in \eqref{s_1}, we utilize Quadratic Transform \cite{9293031} to reformulate the sum-of-ratio problems, 
 \begin{subequations}
  \begin{align}
   P_6:~&\mathop {\max }\limits_{\boldsymbol{\eta},\boldsymbol{\varpi}}~s_2(\boldsymbol{\eta},\boldsymbol{\gamma},\boldsymbol{\varpi},\vartheta)
   \\
   & \text{subject~to} \nonumber \\
   & \eta_k\geq\frac{(2^{S_\text{min}}-1){B_{k}(\boldsymbol{\eta})}}{p_u{\bar{A}_{k}(\boldsymbol{\eta})}}=\eta_{k,\text{min}},~\forall k,
   \\
   &
   0\leq \eta_k\leq 1,~\forall k,
  \end{align}
  \label{Uplink_optimization_P4}
 \end{subequations}
where $\bar{A}_{k}(\boldsymbol{\eta})=\big{|} \displaystyle\sum\nolimits_{m=1}^M \zeta_m \textbf{v}_{mk}{\textbf{h}}_{mk}\big{|}^2,~\forall k$, and $s_2(\boldsymbol{\eta},\boldsymbol{\gamma},\boldsymbol{\varpi},\vartheta)$ is the new objective function expressed as \eqref{s_2_uplink} at the top of the next page, 
\begin{figure*}[t!]
\begin{equation}
\begin{array}{ll}
   \displaystyle s_2(\boldsymbol{\eta},\boldsymbol{\gamma},\boldsymbol{\varpi},\vartheta)&\displaystyle=\sum\nolimits_{k=1}^K\left(\text{log}(1+\gamma_{k})-\gamma_{k}\right)+\sum\nolimits_{k=1}^K\left(2\varpi_{k}\sqrt{(1+\gamma_{k})A_{k}(\boldsymbol{\eta})}-\varpi_{k}^2\left(A_{k}(\boldsymbol{\eta})+B_{k}(\boldsymbol{\eta})\right)\right)-\vartheta\bar{P}_\text{tot}\left(\boldsymbol{\eta}\right)\\&\displaystyle=\sum\nolimits_{k=1}^K\left(\text{log}(1+\gamma_{k})-\gamma_{k}\right)+\sum\nolimits_{k=1}^K\left(2\varpi_{k}\sqrt{(1+\gamma_{k})p_u\eta_k}\big{|} \sum\nolimits_{m=1}^M \zeta_m \textbf{v}_{mk}\textbf{h}_{mk}\big{|}-\varpi_{k}^2\left(A_{k}(\boldsymbol{\eta})+B_{k}(\boldsymbol{\eta})\right)\right)-\vartheta\bar{P}_\text{tot}\left(\boldsymbol{\eta}\right).
\end{array}\label{s_2_uplink}
\vspace{-6 pt}
  \end{equation}
  \vspace{-15pt}
  \hrulefill
  \end{figure*}
in which $\boldsymbol{\varpi}=[\varpi_{1},...,\varpi_{K}]^T\in\mathbb{R}^{K\times 1}$ is a new set of auxiliary variables with $\boldsymbol{\gamma}$ kept constant as $\gamma_{k}^*=\frac{\displaystyle A_{k}(\boldsymbol{\eta})}{\displaystyle B_{k}(\boldsymbol{\eta})},~\forall k$. Similarly, when we
hold $\boldsymbol{\eta}$ fixed, the optimal auxiliary variable in $\boldsymbol{\varpi}$ that maximizes the objective
function $s_2(\boldsymbol{\eta},\boldsymbol{\gamma},\boldsymbol{\varpi},\vartheta)$ can be obtained by taking the
first-order derivative of \eqref{s_2_uplink} and setting it to zero, namely, $\partial s_2(\boldsymbol{\eta},\boldsymbol{\gamma},\boldsymbol{\varpi},\vartheta)/{\partial \varpi_{k}}=0$. Then, we can have the optimum of $\varpi_{k}^*=\frac{ \displaystyle\sqrt{(1+\gamma_{k})p_u\eta_k}\big{|} \sum\nolimits_{m=1}^M \zeta_m \textbf{v}_{mk}\textbf{h}_{mk}\left(\textbf{u}_k\right)\big{|}}{\displaystyle A_{k}(\boldsymbol{\eta})+B_{k}(\boldsymbol{\eta})},~\forall k$. When $\boldsymbol{\gamma}$ and $\boldsymbol{\varpi}$ are held constant, the optimal value of $\boldsymbol{\eta}$ is obtained as \eqref{eta_optimal_EE} at the top of this page by solving $\partial s_2(\boldsymbol{\eta},\boldsymbol{\gamma},\boldsymbol{\varpi},\vartheta)/{\partial \eta_{k}}=0$, with $\bar{A}_{k}(\boldsymbol{\eta})>0,~\forall k$\cite{10256066,10812717,10878991}. The proposed FP algorithm for EE maximization can be stated in Algorithm \ref{Algorithm_FP} to deliver the optimized power control coefficients.
\begin{figure*}[t!]
\begin{equation}
\begin{array}{ll}
  \displaystyle \eta_k^*=\text{min}\left\{
   1,\text{max}\left\{\eta_{k,\text{min}},\frac{\displaystyle(1+\gamma_k)\varpi_k^2\big{|} \sum\nolimits_{m=1}^M \zeta_m \textbf{v}_{mk}\textbf{h}_{mk}\left(\textbf{u}_k\right)\big{|}^2}{\displaystyle
p_u\left[
\sum\nolimits_{k'=1}^K \varpi_{k'}^2\left(\big{|} \sum\nolimits_{m=1}^M \zeta_m \textbf{v}_{mk'}\textbf{h}_{mk}\big{|}^2
+\sum\nolimits_{m=1}^M\zeta_m (1-\zeta_m )\textbf{v}_{mk'}\text{diag}\left(\textbf{h}_{mk}\left(\textbf{u}_k\right)\textbf{h}_{mk}\left(\textbf{u}_k\right)^H\right)\textbf{v}_{mk'}^H\right)+\frac{\displaystyle\vartheta}{\displaystyle\varepsilon_k}
\right]^2}\right\}
   \right\},~\forall k.
\end{array}\label{eta_optimal_EE}
\vspace{-5 pt}
  \end{equation}
  \vspace{-10pt}
  \hrulefill
  \end{figure*}
  \begin{algorithm}[t!]
   \caption{Power Control to Maximize EE}
\begin{algorithmic}[1]
\renewcommand{\algorithmicrequire}{\textbf{Inputs:}}
\Require
$\epsilon>0$ (tolerance), iteration index $i\gets 0$; 
\State Initialize $\eta_{k}^{0},~\forall k$.
\Repeat
\State Update $\vartheta^{(i+1)}=\frac{\displaystyle\sum\nolimits_{k=1}^K\text{log}_2\left(1+\frac{{A_{k}(\boldsymbol{\eta}^{(i)})}}{{B_{k}(\boldsymbol{\eta}^{(i)})}}\right)}{\displaystyle\bar{P}_\text{tot}\left(\boldsymbol{\eta}^{(i)}\right)}$.
\State Update $\eta_{k,\text{min}}^{(i+1)}=\frac{\displaystyle (2^{S_\text{min}}-1){B_{k}(\boldsymbol{\eta}^{(i)})}}{\displaystyle p_u{\bar{A}_{k}(\boldsymbol{\eta}^{(i)})}},~\forall k$.
\State Update $\gamma_{k}^{(i+1)}=\frac{\displaystyle A_{k}(\boldsymbol{\eta}^{(i)})}{\displaystyle B_{k}(\boldsymbol{\eta}^{(i)})},~\forall k$.
\State Update $\varpi_{k}^{(i+1)}=\frac{\displaystyle \sqrt{(1+\gamma_{k}^{(i+1)})A_{k}(\boldsymbol{\eta}^{(i)})}}{\displaystyle A_{k}(\boldsymbol{\eta}^{(i)})+B_{k}(\boldsymbol{\eta}^{(i)})},~\forall k$.
\State  Update $\eta_{k}^{(i+1)}$, $~\forall k$ with $\vartheta^{(i+1)}$, $\eta_{k,\text{min}}^{(i+1)}$, $\gamma_{k}^{(i+1)}$ and $\varpi_{k}^{(i+1)}$ by \eqref{eta_optimal_EE}.
\State Update $\displaystyle\text{SE}_\text{sum}^{(i+1)}=\sum\nolimits_{k=1}^K\text{SE}_k\left(\boldsymbol{\eta}^{(i+1)}\right).$
\State Update $\displaystyle \text{EE}^{(i+1)}=B\cdot\text{SE}_\text{sum}^{(i+1)}/P_\text{tot}^{(i+1)}$.
\State $i\gets i+1$
\Until 
$\big{|}\text{EE}^{(i)}-\text{EE}^{(i-1)}\big{|}^2\leq \epsilon$ is satisfied.
\renewcommand{\algorithmicensure}{\textbf{Output:}}
\Ensure
 $\eta_{k}^{\text{opt}}=\eta_{k}^{(i)},~\forall k$.
\end{algorithmic}
\label{Algorithm_FP}
 \end{algorithm}
\vspace{-10pt}
  \subsection{FAS Position Selection}
Note that $\bar{P}_\text{tot}$ in \eqref{EE_optimization_Dink} is independent of the FAS position. Thus, for given $\eta_{k},~b_m,~\forall k,m$, the subproblem of the FAS position selection is formulated as 
\begin{subequations}
\begin{align}
  P_7:~ &\mathop {\max }\limits_{\textbf{u}}~\sum\nolimits_{k=1}^K\text{SE}_k\left(\textbf{u}\right)\\
   & \text{subject~to}~~
   \eqref{SE_cons},\eqref{FAS_position_cons}.
  \end{align}
  \label{EE_optimization_P5}
 \end{subequations}
where $\textbf{u}=[\textbf{u}_1;...;\textbf{u}_K]\in\mathbb{C}^{2K\times 1}$. Moreover, the quality-of-service (Qos) constraint in \eqref{SE_cons}, $\text{SE}_k\left(\textbf{u}\right)\geq S_\text{min},~\forall k$, by re-formulating \eqref{A_k}-\eqref{B_k}, can be re-written as
\begin{equation}
    \begin{array}{cc}
g_k(\textbf{u})\triangleq\left(2^{S_\text{min}}-1\right)B_k(\textbf{u})-A_k(\textbf{u}).
\end{array}
\end{equation}
Then, for each QoS constraint, we can introduce the relevant quadratic loss function \cite{9709200}
\begin{equation}
\psi_k(\textbf{u})=\left[\text{max}\left(0,g_k(\textbf{u})\right)\right]^2,~\forall k.
\end{equation}
Note that $g_k(\textbf{u})$ is convex and $\psi_k(\textbf{u})$ is smooth. Then, for a given penalty coefficient $\xi$, the penalized objective function of $P_7$ can be given by
\begin{equation}
    \begin{array}{cc}
   \displaystyle  f_\xi(\textbf{u})=f(\textbf{u})-\xi\sum\nolimits_{k=1}^K\psi_k(\textbf{u}),
    \end{array}
\end{equation}
where $f(\textbf{u})=\sum\nolimits_{k=1}^K\text{SE}_k(\textbf{u})$. We remark that the above regularized objective is formed in the context of maximization. Also note that the value of 
the penalty coefficient $\xi$ should be selected appropriately \cite{9709200}. Then, the key to the penalty method is to solve
the following regularized optimization problem for a given $\xi$
\begin{subequations}
\begin{align}
  P_8:~ &\mathop {\max }\limits_{\textbf{u}}~ f_\xi(\textbf{u})\\
   & \text{subject~to} \nonumber \\
   &\textbf{u}_k\in \mathcal{C}_k,~\forall k.
   \label{position_cons}
  \end{align}
  \label{f_xi}
 \end{subequations}
 We are now in a position to apply the APGA algorithm to solve \eqref{f_xi}. To implement the proposed APGA algorithm, we first need to compute $\displaystyle 
\nabla_{\textbf{u}} f_\xi(\textbf{u})$, which is given by
\begin{equation}
\begin{array}{ll}
   \displaystyle 
   \nabla_{\textbf{u}} f_\xi(\textbf{u}) &\displaystyle=   \nabla_{\textbf{u}} f(\textbf{u})-\xi\sum\nolimits_{k=1}^K\nabla_{\textbf{u}}\left[\text{max}\left(0,g_k(\textbf{u})\right)\right]^2\\&\displaystyle=\nabla_{\textbf{u}} f(\textbf{u})-\xi\sum\nolimits_{k=1}^K2\left[\text{max}\left(0,g_k(\textbf{u})\right)\right]\nabla_{\textbf{u}}g_k(\textbf{u}),
\end{array}
  \end{equation}
where
  \begin{equation}
\begin{array}{ll}
   \displaystyle 
   \nabla_{\textbf{u}} f(\textbf{u})=\left[\frac{\partial f(\textbf{u}) }{\partial\textbf{u}_1};\frac{\partial f(\textbf{u}) }{\partial\textbf{u}_2};...;\frac{\partial f(\textbf{u}) }{\partial\textbf{u}_K}\right]\in\mathbb{C}^{2K\times1},
\end{array}\label{f_gradient}
  \end{equation}
   \begin{equation}
\begin{array}{ll}
   \displaystyle 
   \nabla_{\textbf{u}} g_k(\textbf{u})=\left[\frac{\partial g_k(\textbf{u}) }{\partial\textbf{u}_1};\frac{\partial g_k(\textbf{u}) }{\partial\textbf{u}_2};...;\frac{\partial g_k(\textbf{u}) }{\partial\textbf{u}_K}\right]\in\mathbb{C}^{2K\times1},~\forall k.
\end{array}\label{g_gradient}
  \end{equation}
  Based on \eqref{rho_mk}-\eqref{f_mk}, we can first compute
$\frac{\displaystyle\partial\textbf{f}_{mk}(\textbf{u}_k)}{\displaystyle\partial\textbf{u}_k}$. Accordingly,
$\left[\frac{\displaystyle\partial\textbf{f}_{mk}(\textbf{u}_k)}{\displaystyle\partial\textbf{u}_k}\right]_1$ and $\left[\frac{\displaystyle\partial\textbf{f}_{mk}(\textbf{u}_k)}{\displaystyle\partial\textbf{u}_k}\right]_2$ represent the elements of the first column and the second column of the gradient matrix of $\textbf{f}_{mk}(\textbf{u}_k)$, respectively, which are given by
 \begin{equation}
\begin{array}{ll}
\left[\frac{\displaystyle\partial\textbf{f}_{mk}(\textbf{u}_k)}{\displaystyle\partial\textbf{u}_k}\right]_1&\displaystyle=\frac{\displaystyle\partial\textbf{f}_{mk}(\textbf{u}_k)}{\displaystyle\partial x_k}\\&\displaystyle=j\frac{\displaystyle 2\pi}{\displaystyle \lambda}\text{diag}\left(\sin \theta^t_{mk,l}\cos \phi^t_{mk,l}\right)_1^{L_{mk,t}}\textbf{f}_{mk}(\textbf{u}_k),
   \end{array}
  \end{equation} 
   \begin{equation}
\begin{array}{ll}
\left[\frac{\displaystyle\partial\textbf{f}_{mk}(\textbf{u}_k)}{\displaystyle\partial\textbf{u}_k}\right]_2\displaystyle=\frac{\displaystyle\partial\textbf{f}_{mk}(\textbf{u}_k)}{\displaystyle\partial y_k}\displaystyle=j\frac{\displaystyle 2\pi}{\displaystyle \lambda}\text{diag}\left(\cos \theta^t_{mk,l}\right)_1^{L_{mk,t}}\textbf{f}_{mk}(\textbf{u}_k),
   \end{array}
  \end{equation} 
where $\text{diag}(a_l)_1^L$ represents a $L\times L$ diagonal matrix, which $l$-th element is $a_l$. Then, to deliver the proposed gradients,
we first define $\textbf{f}_{kk'}(\textbf{u})=\sqrt{\eta_{k'}}
  \displaystyle \sum\nolimits_{m=1}^M \zeta_m\textbf{v}_{mk}{\textbf{h}}_{mk'}\left(\textbf{u}_{k'}\right)$. Referring to \cite{11196010,9217298} and based on the chain rule, we can calculate $\frac{\displaystyle\partial f(\textbf{u}) }{\displaystyle\partial\textbf{u}_i}$ as \eqref{g_gradient_per} and $\frac{\displaystyle\partial g_k(\textbf{u}) }{\displaystyle\partial\textbf{u}_i},~\forall k$ as \eqref{g_k_gradient_per} at top of this page,
  \begin{figure*}
 \begin{equation}
\begin{array}{ll}
   \displaystyle 
   \frac{\partial f(\textbf{u}) }{\partial\textbf{u}_i}&\displaystyle=\frac{\partial}{\partial\textbf{u}_i}\sum\nolimits_{k=1}^K \text{log}_2\left(1+\frac{\displaystyle 
   p_u\big{|} \textbf{f}_{kk}(\textbf{u}) \big{|}^2
 }{\displaystyle
   p_u\sum\nolimits_{k'\neq k}\big{|}\textbf{f}_{kk'}(\textbf{u})\big{|}^2+\sigma^2 \displaystyle \sum\nolimits_{m=1}^M\big{|}  \zeta_m \textbf{v}_{mk}\big{|}^2+  \displaystyle\sum\nolimits_{m=1}^M \textbf{v}_{mk}\textbf{R}_{\tilde{\textbf{w}}_{m}}(\textbf{u})
  \textbf{v}_{mk}^H
   }\right)\\&\displaystyle=\frac{1}{\text{ln}2}\left(\begin{array}{ll}\displaystyle
   \sum\nolimits_{k=1}^K
   
    \frac{
\frac{\displaystyle\partial}{\displaystyle\partial\textbf{u}_i}\left(\displaystyle
p_u\sum\nolimits_{k'=1}^K\big{|}\textbf{f}_{kk'}(\textbf{u})\big{|}^2+\sigma^2 \displaystyle \sum\nolimits_{m=1}^M\big{|}  \zeta_m \textbf{v}_{mk}\big{|}^2+  \displaystyle\sum\nolimits_{m=1}^M \textbf{v}_{mk}\textbf{R}_{\tilde{\textbf{w}}_{m}}(\textbf{u})
  \textbf{v}_{mk}^H\right)
  }{\displaystyle
p_u\sum\nolimits_{k'=1}^K\big{|}\textbf{f}_{kk'}(\textbf{u})\big{|}^2+\sigma^2 \displaystyle \sum\nolimits_{m=1}^M\big{|}  \zeta_m \textbf{v}_{mk}\big{|}^2+  \displaystyle\sum\nolimits_{m=1}^M \textbf{v}_{mk}\textbf{R}_{\tilde{\textbf{w}}_{m}}(\textbf{u})
  \textbf{v}_{mk}^H}
\\\displaystyle-\sum\nolimits_{k=1}^K
    \frac{
\frac{\displaystyle\partial}{\displaystyle\partial\textbf{u}_i}\left(\displaystyle
p_u\sum\nolimits_{k'\neq k}^K\big{|}\textbf{f}_{kk'}(\textbf{u})\big{|}^2+\sigma^2 \displaystyle \sum\nolimits_{m=1}^M\big{|}  \zeta_m \textbf{v}_{mk}\big{|}^2+  \displaystyle\sum\nolimits_{m=1}^M \textbf{v}_{mk}\textbf{R}_{\tilde{\textbf{w}}_{m}}(\textbf{u})
  \textbf{v}_{mk}^H\right)
  }{\displaystyle
p_u\sum\nolimits_{k'\neq k}^K\big{|}\textbf{f}_{kk'}(\textbf{u})\big{|}^2+\sigma^2 \displaystyle \sum\nolimits_{m=1}^M\big{|}  \zeta_m \textbf{v}_{mk}\big{|}^2+  \displaystyle\sum\nolimits_{m=1}^M \textbf{v}_{mk}\textbf{R}_{\tilde{\textbf{w}}_{m}}(\textbf{u})
  \textbf{v}_{mk}^H}\end{array}\right)
,
\end{array}\label{g_gradient_per}
  \end{equation}
   \vspace{-6pt}
  \hrulefill
    \end{figure*}
    \begin{figure*}
 \begin{equation}
\begin{array}{ll}
   \displaystyle 
   \frac{\partial g_k(\textbf{u}) }{\partial\textbf{u}_i}&\displaystyle=\frac{\partial\left(\left(2^{S_\text{min}}-1\right)\left(\displaystyle
   p_u\sum\nolimits_{k'\neq k}\big{|}\textbf{f}_{kk'}(\textbf{u})\big{|}^2+\sigma^2 \displaystyle \sum\nolimits_{m=1}^M\big{|}  \zeta_m \textbf{v}_{mk}\big{|}^2+  \displaystyle\sum\nolimits_{m=1}^M \textbf{v}_{mk}\textbf{R}_{\tilde{\textbf{w}}_{m}}(\textbf{u})
  \textbf{v}_{mk}^H\right)-p_u\big{|} \textbf{f}_{kk}(\textbf{u}) \big{|}^2\right)}{\partial\textbf{u}_i}
   ,
\end{array}\label{g_k_gradient_per}
  \end{equation}
  \vspace{-12pt}
  \hrulefill
    \end{figure*}
  where
 \begin{equation}
\begin{array}{ll}
   \displaystyle\frac{   \displaystyle\partial|\textbf{f}_{kk'}(\textbf{u})|^2}{   \displaystyle\partial\textbf{u}_i}\displaystyle=\frac{   \displaystyle\partial\textbf{f}_{kk'}^H(\textbf{u})\textbf{f}_{kk'}(\textbf{u})}{   \displaystyle\partial\textbf{u}_i}
 \displaystyle=
       2\mathfrak{RE}\left\{\textbf{f}_{kk'}^H(\textbf{u})\frac{   \displaystyle\partial\textbf{f}_{kk'}(\textbf{u})}{   \displaystyle\partial\textbf{u}_i}\right\},
\end{array}\label{g_kk_G}
  \end{equation}
and $\displaystyle\frac{   \displaystyle\partial\textbf{f}_{kk'}(\textbf{u})}{   \displaystyle\partial\textbf{u}_i}$ is given by
  \begin{equation}
\begin{array}{ll}
   \displaystyle\displaystyle\frac{   \displaystyle\partial\textbf{f}_{kk'}(\textbf{u})}{   \displaystyle\partial\textbf{u}_i}&\displaystyle=\sqrt{\eta_{k'}}\sum\nolimits_{m=1}^M\zeta_{m}\frac{   \displaystyle\partial\textbf{v}_{mk}\textbf{G}_{mk'}^H\boldsymbol{\Sigma}_{mk'}\textbf{f}_{mk'}\left(\textbf{u}_{k'}\right)}{   \displaystyle\partial\textbf{u}_i}
     \\&\displaystyle=\left\{\begin{array}{ll}
\displaystyle \sum\nolimits_{m=1}^M\sqrt{\eta_{k'}}\zeta_{m}\frac{   \displaystyle\partial\textbf{f}_{mk'}^H\left(\textbf{u}_{k'}\right)}{   \displaystyle\partial\textbf{u}_{k'}}\boldsymbol{\Sigma}_{mk'}^H\textbf{G}_{mk'}\textbf{v}_{mk}^H
,~i=k'
\\
0,~\text{otherwise}
     \end{array}
     
     \right..
\end{array}\label{g_kk'_gradient}
  \end{equation}
Meanwhile, the gradient of the noise and quantization noise term in \eqref{g_gradient_per} can be given by \eqref{noise_gradient} at the top of this page, where $\textbf{D}_{mk,v}\sim\mathbb{C}^{N\times N}$ is a diagonal matrix with $\left[\textbf{D}_{mk,v}\right]_n=\big{|}\left[\textbf{v}_{mk}\right]_n\big{|}^2,~\forall m,~k$.
\begin{figure*}
  \begin{equation}
      \begin{array}{ll}
 \displaystyle\frac{   \displaystyle\partial\left(\sigma^2 \displaystyle \sum\nolimits_{m=1}^M\big{|}  \zeta_m \textbf{v}_{mk}\big{|}^2+  \displaystyle\sum\nolimits_{m=1}^M \textbf{v}_{mk}\textbf{R}_{\tilde{\textbf{w}}_{m}}(\textbf{u})
  \textbf{v}_{mk}^H\right)}{   \displaystyle\partial\textbf{u}_i}&\displaystyle=\sum\nolimits_{m=1}^Mp_u\zeta_m(1-\zeta_m)\displaystyle\frac{   \displaystyle\partial\left(\displaystyle \textbf{v}_{mk}\text{diag}\left(\sum\nolimits_{k'=1}^K\eta_{k'}{\textbf{h}}_{mk'}\left(\textbf{u}_{k'}\right){\textbf{h}}_{mk'}\left(\textbf{u}_{k'}\right)^H+\sigma^2\textbf{I}_{N}\right)
  \textbf{v}_{mk}^H\right)}{   \displaystyle\partial\textbf{u}_i}\\&\displaystyle=\sum\nolimits_{m=1}^Mp_u\zeta_m(1-\zeta_m)\displaystyle\displaystyle\frac{   \displaystyle\displaystyle \textbf{v}_{mk}\partial\left(\text{diag}\left(\sum\nolimits_{k'=1}^K\eta_{k'}{\textbf{h}}_{mk'}\left(\textbf{u}_{k'}\right){\textbf{h}}_{mk'}\left(\textbf{u}_{k'}\right)^H\right)\right)
  \textbf{v}_{mk}^H}{   \displaystyle\partial\textbf{u}_i}



\\&\displaystyle=\left\{\begin{array}{ll}
\displaystyle 2p_u\sum\nolimits_{m=1}^M\sum\nolimits_{n=1}^N\zeta_m(1-\zeta_m)\displaystyle \eta_{k'}\big{|}\left[\textbf{v}_{mk}\right]_n\big{|}^2\displaystyle\mathfrak{RE}\left\{\left[\textbf{h}_{mk'}\left(\textbf{u}_{k'}\right)\right]_n\frac{   \displaystyle\partial\left[\textbf{h}_{mk'}\left(\textbf{u}_{k'}\right)\right]_n}{   \displaystyle\partial\textbf{u}_i}\right\}
,~i=k'
\\
0,~\text{otherwise}
     \end{array}
     
     \right.
     \\&\displaystyle=\left\{\begin{array}{ll}
\displaystyle 2p_u\sum\nolimits_{m=1}^M\zeta_m(1-\zeta_m)\displaystyle \eta_{k'}\displaystyle\mathfrak{RE}\left\{\frac{   \displaystyle\partial\textbf{f}_{mk'}^H\left(\textbf{u}_{k'}\right)}{   \displaystyle\partial\textbf{u}_{k'}}\boldsymbol{\Sigma}_{mk'}^H\textbf{G}_{mk'}\textbf{D}_{mk,v}\textbf{h}_{mk'}\left(\textbf{u}_{k'}\right)\right\}
,~i=k'
\\
0,~\text{otherwise}
     \end{array}
     
     \right..
      \end{array}
      \label{noise_gradient}
       \vspace{-8pt}
  \end{equation}
  \hrulefill
   \vspace{-10pt}
  \end{figure*} 
  \begin{algorithm}[t!]
   \caption{Backtracking Line Search for Step Size} 
\begin{algorithmic}[1]
\renewcommand{\algorithmicrequire}{\textbf{Inputs:}}
\Require
$\iota\in(0,1)$, $\varrho\in(0,1)$, $\alpha>0$
\renewcommand{\algorithmicensure}{\textbf{Output:}}
\Ensure
 $\alpha_1^{(n+1)},~\alpha_2^{(n+1)}$
\State Initialize the step size as $\alpha_1^{(n+1)}=\alpha$, $\alpha_2^{(n+1)}=\alpha$
\Repeat 
\State $\textbf{v}^{n+1}=P\left(\textbf{x}^n+\alpha^{(n+1)}\nabla_{\textbf{x}^n} f_\xi(\textbf{x}^n)\right)$ 
\State $\alpha_1^{(n+1)}=\iota \alpha_1^{(n+1)}$
\Until {$f_\xi(\textbf{v}^{n+1})\geq f_\xi(\textbf{v}^{n})+\varrho\alpha_1^{(n+1)}||\textbf{v}^{n+1}-\textbf{x}^n||$} 
\Repeat 
\State $\textbf{z}^{n+1}=P\left(\textbf{y}^n+\alpha_2^{(n+1)}\nabla_{\textbf{y}^n}  f_\xi(\textbf{y}^n)\right)$ 
\State $\alpha_2^{(n+1)}=\iota \alpha_2^{(n+1)}$
\Until {$f_\xi(\textbf{z}^{n+1})\geq f_\xi(\textbf{y}^{n})+\varrho\alpha_2^{(n+1)}||\textbf{z}^{n+1}-\textbf{y}^{n}||$}  
\end{algorithmic}
\label{BLS}
 \end{algorithm} 
\subsubsection{Projection on $\mathcal{C}$}
  To guarantee the constraint in \eqref{FAS_position_cons}, each update for FAS positions is followed by a projection operation $\text{P}_{\mathcal{C}}\left(\textbf{u}\right)$, which is defined as
\begin{equation}
  \begin{array}{cc}
[\text{P}_{\mathcal{C}}\left(\textbf{u}_k\right)]_{p}=\displaystyle\left\{\begin{array}{ll}
d_\text{min},~\text{if} ~[\textbf{u}_k]_p<d_\text{min}
\\
{[}\textbf{u}_k{]}_p,~\text{if}~ d_\text{min}<[\textbf{u}_k]_p<d_\text{max}
\\
d_\text{max},~\text{if}~ [\textbf{u}_k]_p>d_\text{max}
\end{array}
\right.,~\forall k,
\end{array}
\end{equation}
where $[\cdot]_p$ denotes the $p$-th entry of the argument. Note that the convergence behavior of the gradient ascent method highly depends on the step size, which can be calculated by the backtracking line search\cite{hu2025uplinktransmissiondesignfluid,10388242}. Specifically, the step size iteratively decreases during the search process with a scaling factor $\iota\in\left(0,1\right)$ and a control parameter $\varrho\in\left(0,1\right)$, until it satisfies the Armijo-Goldstein condition\cite{hu2025uplinktransmissiondesignfluid}.
Then, the step size design for FAS position optimization follows Algorithm \ref{BLS} at the top of this page with $\textbf{x}=\textbf{u}$. Then, we illustrate
the iterative procedure of the APGA algorithm in Algorithm \ref{Algorithm_APG_MA} at the top of this page.
 \begin{algorithm}[t!]
   \caption{FAS Position Optimization to Maximize EE} 
\begin{algorithmic}[1]
\renewcommand{\algorithmicrequire}{\textbf{Inputs:}}
\Require
$\epsilon$ (tolerance), $\text{I}_\text{max}$; $\textbf{u}^0\geq 0$, $t_0=t_1=1$
\renewcommand{\algorithmicensure}{\textbf{Output:}}
\Ensure
$\textbf{u}$
\State Initialize $\textbf{u}^1=\textbf{z}^1=\textbf{u}^0$,
\For{$\text{n}= 1 :\text{I}_\text{max}$}
\State $\textbf{y}^n=\textbf{u}^n+\frac{\displaystyle t_{n-1}}{\displaystyle t_n}(\textbf{z}^n-\textbf{u}^n)+\frac{\displaystyle t_{n-1}-1}{\displaystyle t_n}(\textbf{u}^n-\textbf{u}^{n-1})$
\State Obtain the step size based on Algorithm \ref{BLS}
\State $\textbf{v}^{n+1}=P_{\mathcal{C}}(\textbf{u}^n+\alpha_1^{(n+1)}\nabla f_\xi(\textbf{u}^n))$
\State $\textbf{z}^{n+1}=P_{\mathcal{C}}(\textbf{y}^n+\alpha_2^{(n+1)}\nabla f_\xi(\textbf{y}^n))$
 \State $\textbf{u}^{n+1}=\left\{\begin{array}{ll}
  \textbf{z}^{n+1},~f_\xi(\textbf{z}^{n+1})>f_\xi(\textbf{v}^{n+1})
  \\
  \textbf{v}^{n+1},~\text{otherwise}
 \end{array}\right.$
\State $t_{n+1}=\frac{\displaystyle 1+\sqrt{4t_n^2+1}}{\displaystyle 2}$
\If {$\big{|}f_\xi(\textbf{u}^{n+1})-f_\xi(\textbf{u}^{n})\big{|}\leq \epsilon$}
\State $\textbf{break}$
\EndIf
\EndFor
\end{algorithmic}
\label{Algorithm_APG_MA}
 \end{algorithm}

 \subsection{Bit allocation}
For given $\eta_{k},~\textbf{u}_k,~\forall k$, we first release the constraint for bit allocation $b_m\in\{b_\text{min}, b_\text{max}\},~\forall m$ to 
$b_\text{min} \leq b_m\leq b_\text{max},~\forall m$. Then,
the subproblem of bit allocation is formulated as
\begin{subequations}
  \begin{align}
   P_9:~&\mathop {\max }\limits_{\textbf{b}}~\displaystyle\sum\nolimits_{k=1}^K\text{SE}_k\left(\textbf{b}\right)-\vartheta\bar{P}_\text{tot}\left(\textbf{b}\right)
   \\
   & \text{subject~to} ~~
\eqref{SE_cons},\eqref{bit_cons},
  \end{align}
  \label{EE_optimization_bit}
 \end{subequations}
where $\left(\textbf{b}\right)=[b_1,...,b_M]^T\in\mathbb{C}^{M\times 1}$. Meanwhile, $\vartheta^{*}=\frac{\displaystyle\sum\nolimits_{k=1}^K\text{SE}_k\left(\textbf{b}\right)}{\displaystyle\bar{P}_\text{tot}\left(\textbf{b}\right)}$ when $\textbf{b}$ is held fixed.  Similar to the FAS position optimization, the QoS $\text{SE}_k\left(\textbf{b}\right)\geq S_\text{min},~\forall k$, by utilizing \eqref{A_k}-\eqref{B_k}, can be re-written as
\begin{equation}
    \begin{array}{cc}
g_k(\textbf{b})\triangleq\left(2^{S_\text{min}}-1\right)B_k(\textbf{b})-A_k(\textbf{b}).
\end{array}
\end{equation}
Accordingly, we can introduce the relevant quadratic loss function for each QoS constraint\cite{9709200}
\begin{equation}
\psi_k(\textbf{b})=\left[\text{max}\left(0,g_k(\textbf{b})\right)\right]^2,~\forall k.
\end{equation}
Then, for a given penalty coefficient $\xi$, the penalized objective function of $P_9$ can be given by
\begin{equation}
    \begin{array}{cc}
     f_\xi(\textbf{b})=f(\textbf{b})-\xi\sum\nolimits_{k=1}^K\psi_k(\textbf{b}),
    \end{array}
\end{equation}
where $f(\textbf{b})=\sum\nolimits_{k=1}^K\text{SE}_k(\textbf{b})-\vartheta\bar{P}_\text{tot}\left(\textbf{b}\right)$. The above regularized objective is formed in the context of maximization \cite{9709200}. Then, the key to the penalty method is to solve
the following regularized optimization problem for a given $\xi$
\begin{subequations}
\begin{align}
  P_{10}:~ &\mathop {\max }\limits_{\textbf{b}_k}~ f_\xi(\textbf{b})\\
   & \text{subject~to} \nonumber \\
   &b_\text{min} \leq b_m\leq b_\text{max},~\forall m.
   \label{position_bit}
  \end{align}
  \label{f_bit}
 \end{subequations}
 We now to apply the APGA algorithm to solve \eqref{f_bit} and compute $\displaystyle 
\nabla_{\textbf{b}} f_\xi(\textbf{b})$ with
\begin{equation}
\begin{array}{ll}
   \displaystyle 
   \nabla_{\textbf{b}} f_\xi(\textbf{b}) &\displaystyle=   \nabla_{\textbf{b}} f(\textbf{b})-\xi\sum\nolimits_{k=1}^K\nabla_{\textbf{b}}\left[\text{max}\left(0,g_k(\textbf{b})\right)\right]^2\\&\displaystyle=\nabla_{\textbf{b}} f(\textbf{b})-\xi\sum\nolimits_{k=1}^K2\left[\text{max}\left(0,g_k(\textbf{b})\right)\right]\nabla_{\textbf{b}}g_k(\textbf{b}),
\end{array}
  \end{equation}
where
  \begin{equation}
\begin{array}{ll}
   \displaystyle 
   \nabla_{\textbf{b}} f(\textbf{b})=\left[\frac{\partial f(\textbf{b}) }{\partial b_1};\frac{\partial f(\textbf{b}) }{\partial b_2};...;\frac{\partial f(\textbf{b}) }{\partial b_M}\right]\in\mathbb{C}^{M\times1},
\end{array}\label{f_gradient}
  \end{equation}
   \begin{equation}
\begin{array}{ll}
   \displaystyle 
   \nabla_{\textbf{b}} g_k(\textbf{b})=\left[\frac{\partial g_k(\textbf{b}) }{\partial b_1};\frac{\partial g_k(\textbf{b}) }{\partial b_2};...;\frac{\partial g_k(\textbf{b}) }{\partial b_M}\right]\in\mathbb{C}^{M\times1},~\forall k.
\end{array}\label{g_gradient}
  \end{equation}
  Then, to deliver the proposed gradients,
we first define $\textbf{f}_{kk'}(\textbf{b})=\sqrt{\eta_{k'}}
  \displaystyle \sum\nolimits_{m=1}^M \zeta_m(b_m)\textbf{v}_{mk}{\textbf{h}}_{mk'}$. Then, we can calculate $\frac{\displaystyle\partial f(\textbf{b}) }{\displaystyle\partial b_i}$ as \eqref{g_gradient_bit} and $\frac{\displaystyle\partial g_k(\textbf{b}) }{\displaystyle\partial b_i},~\forall k$ as \eqref{g_k_gradient_bit} at top of the next page,
\begin{figure*}
 \begin{equation}
\begin{array}{ll}
   \displaystyle 
   \frac{\partial f(\textbf{b}) }{\partial b_i}&\displaystyle=\frac{\partial}{\partial b_i}\left(\sum\nolimits_{k=1}^K \text{log}_2\left(1+\frac{\displaystyle 
   p_u\big{|} \textbf{f}_{kk}(\textbf{b}) \big{|}^2
 }{\displaystyle
   p_u\sum\nolimits_{k'\neq k}\big{|}\textbf{f}_{kk'}(\textbf{b})\big{|}^2+\sigma^2 \displaystyle \sum\nolimits_{m=1}^M\big{|}  \zeta_m(b_m) \textbf{v}_{mk}\big{|}^2+  \displaystyle\sum\nolimits_{m=1}^M \textbf{v}_{mk}\textbf{R}_{\tilde{\textbf{w}}_{m}}(\textbf{b})
  \textbf{v}_{mk}^H
   }\right)
   -\vartheta\bar{P}_\text{tot}\left(\textbf{b}\right)
   \right)
   
   \\&\displaystyle=\frac{1}{\text{ln}2}\left(\begin{array}{ll}\displaystyle
   \sum\nolimits_{k=1}^K
   
    \frac{
\frac{\displaystyle\partial}{\displaystyle\partial b_i}\left(\displaystyle
p_u\sum\nolimits_{k'=1}^K\big{|}\textbf{f}_{kk'}(\textbf{b})\big{|}^2+\sigma^2 \displaystyle \sum\nolimits_{m=1}^M\big{|}  \zeta_m(b_m) \textbf{v}_{mk}\big{|}^2+  \displaystyle\sum\nolimits_{m=1}^M \textbf{v}_{mk}\textbf{R}_{\tilde{\textbf{w}}_{m}}(\textbf{b})
  \textbf{v}_{mk}^H\right)
  }{\displaystyle
p_u\sum\nolimits_{k'=1}^K\big{|}\textbf{f}_{kk'}(\textbf{b})\big{|}^2+\sigma^2 \displaystyle \sum\nolimits_{m=1}^M\big{|}  \zeta_m(b_m) \textbf{v}_{mk}\big{|}^2+  \displaystyle\sum\nolimits_{m=1}^M \textbf{v}_{mk}\textbf{R}_{\tilde{\textbf{w}}_{m}}(\textbf{b})
  \textbf{v}_{mk}^H}
\\\displaystyle-\sum\nolimits_{k=1}^K
    \frac{
\frac{\displaystyle\partial}{\displaystyle\partial b_i}\left(\displaystyle
p_u\sum\nolimits_{k'\neq k}^K\big{|}\textbf{f}_{kk'}(\textbf{b})\big{|}^2+\sigma^2 \displaystyle \sum\nolimits_{m=1}^M\big{|}  \zeta_m(b_m) \textbf{v}_{mk}\big{|}^2+  \displaystyle\sum\nolimits_{m=1}^M \textbf{v}_{mk}\textbf{R}_{\tilde{\textbf{w}}_{m}}(\textbf{b})
  \textbf{v}_{mk}^H\right)
  }{\displaystyle
p_u\sum\nolimits_{k'\neq k}^K\big{|}\textbf{f}_{kk'}(\textbf{b})\big{|}^2+\sigma^2 \displaystyle \sum\nolimits_{m=1}^M\big{|}  \zeta_m(b_m) \textbf{v}_{mk}\big{|}^2+  \displaystyle\sum\nolimits_{m=1}^M \textbf{v}_{mk}\textbf{R}_{\tilde{\textbf{w}}_{m}}(\textbf{b})
  \textbf{v}_{mk}^H}\end{array}\right)-2N\vartheta\frac{\displaystyle\partial}{\displaystyle\partial b_i}\sum\nolimits_{m=1}^MP_{m,tc}(b_m)
,
\end{array}\label{g_gradient_bit}
 \vspace{-2pt}
  \end{equation}
      \vspace{-10pt}
  \hrulefill
    \end{figure*}
\begin{figure*}
 \begin{equation}
\begin{array}{ll}
   \displaystyle 
   \frac{\partial g_k(\textbf{b}) }{\partial b_i}&\displaystyle=\frac{\partial\left(\left(2^{S_\text{min}}-1\right)\left(\displaystyle
   p_u\sum\nolimits_{k'\neq k}\big{|}\textbf{f}_{kk'}(\textbf{b})\big{|}^2+\sigma^2 \displaystyle \sum\nolimits_{m=1}^M\big{|}  \zeta_m(b_m) \textbf{v}_{mk}\big{|}^2+  \displaystyle\sum\nolimits_{m=1}^M \textbf{v}_{mk}\textbf{R}_{\tilde{\textbf{w}}_{m}}(\textbf{b})
  \textbf{v}_{mk}^H\right)-p_u\big{|} \textbf{f}_{kk}(\textbf{b}) \big{|}^2\right)}{\partial b_i}
   ,
\end{array}\label{g_k_gradient_bit}
  \end{equation}
     \vspace{-10pt}
  \hrulefill
    \end{figure*}
    where
 \begin{equation}
\begin{array}{ll}
   \displaystyle\frac{   \displaystyle\partial|\textbf{f}_{kk'}(\textbf{b})|^2}{   \displaystyle b_i}\displaystyle=\frac{   \displaystyle\partial\textbf{f}_{kk'}^H(\textbf{b})\textbf{f}_{kk'}(\textbf{b})}{   \displaystyle b_i}
\displaystyle=
       2\mathfrak{RE}\left\{\textbf{f}_{kk'}^H(\textbf{b})\frac{   \displaystyle\partial\textbf{f}_{kk'}(\textbf{b})}{   \displaystyle\partial b_i}\right\},
\end{array}\label{g_kk_G_bit}
  \end{equation}
and $\displaystyle\frac{   \displaystyle\partial\textbf{f}_{kk'}(\textbf{b})}{   \displaystyle\partial b_i}$ is given by   \footnote{The exact value of $\zeta_m, \forall m,$ for 1-5 bits can be found in Table I \cite{10858168,10878991}. The purpose of using this simple relationship $\zeta_m\approx1-\frac{\pi\sqrt{3}}{2}2^{-2b_m},~\forall m$, is to facilitate the design of the ADC bits allocation scheme in Section IV. Also, the exact values are used in simulations with equal bit allocation\cite{8756265,10878991}.}
  \begin{equation}
\begin{array}{ll}
   \displaystyle\displaystyle\frac{   \displaystyle\partial\textbf{f}_{kk'}(\textbf{b})}{   \displaystyle\partial b_i}&\displaystyle=\sqrt{\eta_{k'}}\frac{   \displaystyle\partial}{   \displaystyle\partial b_i}\sum\nolimits_{m=1}^M\zeta_{m}\left(b_m\right)\textbf{v}_{mk}\textbf{h}_{mk'}^H
     \\&\displaystyle=
\displaystyle 
\frac{\pi\sqrt{3}}{2}\sqrt{\eta_{k'}}\text{ln}\left(4\right)2^{-2b_i}\textbf{v}_{ik}\textbf{h}_{ik'}^H.
\end{array}\label{g_kk'_gradient_bit}
  \end{equation}
Then, the gradient of the noise and power terms in \eqref{g_gradient_bit} can be given by
  \begin{equation}
      \begin{array}{ll}
 \displaystyle\frac{   \displaystyle\partial\sigma^2 \displaystyle \sum\nolimits_{m=1}^M\big{|}  \zeta_m(b_m) \textbf{v}_{mk}\big{|}^2}{   \displaystyle\partial b_i}&\displaystyle=\sigma^2\displaystyle\frac{   \displaystyle\partial \displaystyle }{   \displaystyle\partial b_i}\sum\nolimits_{m=1}^M\zeta_m(b_m) ^2\big{|}\textbf{v}_{mk}\big{|}^2\\&\displaystyle=
\pi\sqrt{3}\sigma^2\text{ln}\left(4\right)2^{-2b_i}\zeta_i(b_i) \big{|}\textbf{v}_{ik}\big{|}^2,
\end{array}
\end{equation}
 \begin{equation}
      \begin{array}{ll}
 \displaystyle\frac{ \displaystyle\partial\displaystyle \sum\nolimits_{m=1}^MP_{m,tc}(b_m)}{   \displaystyle\partial b_i}&\displaystyle=\displaystyle\frac{   \displaystyle\partial\displaystyle }{   \displaystyle\partial b_i}\sum\nolimits_{m=1}^M\left(c_m(b_m)P_{m,AGC}+2^{b_m}\cdot\text{FOM}_\textit{W}\cdot f_s\right)\\&\displaystyle=
\displaystyle \min\left(b_i-1,1\right)P_{i,AGC}+
\text{ln}\left(2\right)2^{b_i}\text{FOM}_\textit{W}\cdot f_s.
\end{array}
\end{equation}
\begin{figure*}
  \begin{equation}
      \begin{array}{ll}
 \displaystyle\frac{   \displaystyle\partial \displaystyle\sum\nolimits_{m=1}^M \textbf{v}_{mk}\textbf{R}_{\tilde{\textbf{w}}_{m}}(\textbf{b})
  \textbf{v}_{mk}^H }{   \displaystyle\partial b_i}&\displaystyle=\displaystyle\frac{   \displaystyle\partial \displaystyle }{   \displaystyle\partial b_i}
\sum\nolimits_{m=1}^M \textbf{v}_{mk}\zeta_m(b_m)\left(1-\zeta_m(b_m)\right)\text{diag}\left(p_u\sum\nolimits_{k=1}^K\eta_k{\textbf{h}}_{mk}{\textbf{h}}_{mk}^H+\sigma^2\textbf{I}_{N}\right)
  \textbf{v}_{mk}^H
\\&\displaystyle=
\displaystyle
\frac{\pi\sqrt{3}}{2}p_u\left(1-2\zeta_i(b_i)\right)\text{ln}\left(4\right)2^{-2b_i}\textbf{v}_{ik}\text{diag}\left(p_u\sum\nolimits_{k=1}^K\eta_k{\textbf{h}}_{ik}{\textbf{h}}_{ik}^H+\sigma^2\textbf{I}_{N}\right)
  \textbf{v}_{ik}^H.
\end{array}
\label{quanti_bit}
\vspace{-2pt}
\end{equation}
\hrulefill
    \vspace{-10pt}
\end{figure*}
Meanwhile, the gradient of the quantization noise in \eqref{g_gradient_bit} can be given by \eqref{quanti_bit} at the top of the next page.
\subsubsection{Projection}
  To guarantee the constraint in \eqref{bit_cons} and recover the released values as integers, each update for bit allocation is followed by a projection operation $\text{P}_{\mathcal{S}}\left(\textbf{b}\right)$, which is defined as
\begin{equation}
  \begin{array}{cc}
\text{P}_{\mathcal{S}}\left( b_m\right)=\displaystyle\left\{\begin{array}{ll}
b_\text{min},~\text{if} ~b_m\leq b_\text{min}
\\
\text{round}(b_m),~\text{if}~ b_\text{min}<b_m<b_\text{max}
\\
b_\text{max},~\text{if}~ b_m\geq b_\text{max}
\end{array}
\right.,~\forall m.
\end{array}
\end{equation}
Similar to FAS position optimization, the convergence behavior of the gradient ascent method for bit allocation also highly depends on the step size, which can be calculated by the backtracking line search following Algorithm \ref{BLS} with $\textbf{x}=\textbf{b}$\cite{hu2025uplinktransmissiondesignfluid,10388242}. Then, we illustrate
the iterative procedure of the APGA algorithm in Algorithm \ref{Algorithm_APG_bit}.

 \begin{algorithm}[t!]
   \caption{Bit Allocation to Maximize EE} 
\begin{algorithmic}[1]
\renewcommand{\algorithmicrequire}{\textbf{Inputs:}}
\Require
$\epsilon$ (tolerance), $\text{I}_\text{max}$; $\textbf{b}^0\geq 0$, $t_0=t_1=1$
\renewcommand{\algorithmicensure}{\textbf{Output:}}
\Ensure
$\textbf{b}$
\State Initialize $\textbf{b}^1=\textbf{z}^1=\textbf{b}^0$,
\For{$\text{n}= 1 :\text{$\text{I}_\text{max}$}$}
\State Update $\vartheta^{n}=\frac{\displaystyle\sum\nolimits_{k=1}^K\text{log}_2\left(1+\frac{{A_{k}(\textbf{b}^{n})}}{{B_{k}(\textbf{b}^{n})}}\right)}{\displaystyle\bar{P}_\text{tot}\left(\textbf{b}^{n}\right)}$.
\State $\textbf{y}^n=\textbf{b}^n+\frac{\displaystyle t_{n-1}}{\displaystyle t_n}(\textbf{z}^n-\textbf{b}^n)+\frac{\displaystyle t_{n-1}-1}{\displaystyle t_n}(\textbf{b}^n-\textbf{b}^{n-1})$
\State Obtain the step size based on Algorithm \ref{BLS}
\State $\textbf{v}^{n+1}=P_{\mathcal{S}}(\textbf{b}^n+\alpha_1^{(n+1)}\nabla f_\xi(\textbf{b}^n))$
\State $\textbf{z}^{n+1}=P_{\mathcal{S}}(\textbf{y}^n+\alpha_2^{(n+1)}\nabla f_\xi(\textbf{y}^n))$
 \State $\textbf{b}^{n+1}=\left\{\begin{array}{ll}
  \textbf{z}^{n+1},~f_\xi(\textbf{z}^{n+1})>f_\xi(\textbf{v}^{n+1})
  \\
  \textbf{v}^{n+1},~\text{otherwise}
 \end{array}\right.$
\State $t_{n+1}=\frac{\displaystyle 1+\sqrt{4t_n^2+1}}{\displaystyle 2}$
\If {$\big{|}f_\xi(\textbf{b}^{n+1})-f_\xi(\textbf{b}^{n})\big{|}\leq \epsilon$}
\State $\textbf{break}$
\EndIf
\EndFor
\end{algorithmic}
\label{Algorithm_APG_bit}
 \end{algorithm}
 
 \subsection{Alternating Optimization-based Algorithm} 
\label{complexity}
\begin{algorithm}[t!]
   \caption{AO-based algorithm to solve $P_1$ in \eqref{EE_optimization_all}} 
\begin{algorithmic}[1]
\renewcommand{\algorithmicrequire}{\textbf{Inputs:}}
\Require
$\epsilon$ (tolerance), $\boldsymbol{\eta}^0\geq 0$, $\textbf{u}^0\geq 0$, $\textbf{b}^0\geq b_\text{min}$, $\alpha>0$, $\delta>0$, $\text{I}_\text{max}$ (maximum
number of iterations), iteration index $i\gets 0$;
\Repeat
\State Update $\boldsymbol{\eta}^{(i+1)}$ by Algorithm \ref{Algorithm_FP}.
\State Update $\textbf{u}^{(i+1)}$ by Algorithm \ref{Algorithm_APG_MA}.
\State Update $\textbf{b}^{(i+1)}$ by Algorithm \ref{Algorithm_APG_bit}.
\State Update $\textbf{v}_{mk}^{(i+1)},~\forall m, k,$ by \eqref{MMSE}.
\State Update $\text{EE}^{(i+1)},~\forall k$ with $\eta_{k}^{(i+1)}$, $\textbf{u}_k^{(i+1)}$, $b_m^{(i+1)}$, $\forall m,k$.
\State $i\gets i+1$
\Until $\big{|} \text{EE}(\boldsymbol{\eta}^{(i)},\textbf{u}^{(i)},\textbf{b}^{(i)})-\text{EE}(\boldsymbol{\eta}^{(i-1)},\textbf{u}^{(i-1)},\textbf{b}^{(i-1)})\big{|}^2\leq \epsilon$ is satisfied or $i>\text{I}_\text{max}$.
\renewcommand{\algorithmicensure}{\textbf{Output:}}
\Ensure
$\boldsymbol{\eta}^\text{opt}=\boldsymbol{\eta}^{(i)}$, $\textbf{u}^\text{opt}=\textbf{u}^{(i)}$, $\textbf{b}^\text{opt}=\textbf{b}^{(i)}$.
\end{algorithmic}
\label{Algorithm_APG_All}
 \end{algorithm}
In this section, we propose the implementation of the AO-based algorithm to iteratively determine $\eta_{k}$, $\textbf{u}_k$ and $b_m$ $\forall m, k$ for EE maximization, solving $P_1$ in \eqref{EE_optimization_all}. The process is summarized in Algorithm \ref{Algorithm_APG_All}.
Specifically, we randomly initialize the state
and alternately perform Algorithms \ref{Algorithm_FP}, \ref{Algorithm_APG_MA} and \ref{Algorithm_APG_bit}
to increase the EE. Since $\text{EE}(\boldsymbol{\eta}, \textbf{u}, \textbf{b})$ is
non-decreasing over iterations and has an upper bound, Algorithm \ref{Algorithm_APG_All} is guaranteed to converge. Accordingly, the total complexity of Algorithm \ref{Algorithm_APG_All} for maximizing the EE
is bounded by
the complexity required to solve the aforementioned subproblems over $T_{outer}$ outer iterations. Specifically, the complexity of the power control in Algorithm \ref{Algorithm_FP} is  $\mathcal{O}\left(T_{1}MK^2+MK^2N\right)$, where $T_{1}$ is the number of iterations of Algorithm \ref{Algorithm_FP} and $\mathcal{O}\left(MK^2N\right)$ is the complexity computing and storing $\textbf{v}_{mk}{\textbf{h}}_{mk'},~\forall m,~k,$ outside the loop. For the FAS position optimization, the computation is dominated by the gradient evaluations, taking $\mathcal{O}\left(T_2T_{inner,\bf{u}}MK^2L+MK^2NL\right)$, where $L$ is the number of receive paths with $L_{mk,r}=L,~\forall m,k$, $T_2$ is the number of iterations of Algorithm \ref{Algorithm_APG_MA} and $T_{inner,\bf{u}}$ is the average number of backtracking steps per iteration. For the bit allocation, the complexity to solve \ref{Algorithm_APG_bit} is $\mathcal{O}\left(T_3T_{inner,\bf{b}}MK^2+MK^2N\right)$, where $T_3$ is the number of iterations of Algorithm \ref{Algorithm_APG_bit} and $T_{inner,\bf{b}}$ is the average number of backtracking steps per iteration. Notably, updating the local MMSE precoding vectors in step 5 of Algorithm \ref{Algorithm_APG_All} involves $N\times N$ matrix inversions at $M$ APs, introducing a complexity of $\mathcal{O}(MN^3+MKN^2)$. Thus, the total complexity of the proposed AO-based algorithm is roughly given by
$\mathcal{O}\left(T_{outer}\left(\left(T_{1}+T_{2}T_{inner,\bf{u}}L+T_{3}T_{inner,\bf{b}}\right)MK^2+MNK^2L+MN^3\right)\right)$, where $T_{outer}$ is the number of iterations to complete Algorithm \ref{Algorithm_APG_All}. While the proposed AO-based algorithm provides fundamental insights and performance bounds for FAS-enabled cell-free massive MIMO systems, we find that its computational complexity scales rapidly as $M$ and $K$ increase. For ultra-dense 6G networks, maintaining practical feasibility and scalability is of great importance. Thus, to translate the substantial EE gains into real-world deployments, it is imperative to integrate the proposed framework with scalable, low-complexity architectures. Adopting techniques such as user-centric association and partial AP participation will be the next essential steps in our future work to reduce the computational burden while maintaining practical feasibility.

\section{Numerical Results}

In this work, we utilize a three-slope propagation model \cite{10032129,7827017}, where APs and users are uniformly and independently distributed within a 200 m $\times$ 200 m region.
We assume the response matrix as diagonal with independent and identically distributed (i.i.d.) circularly symmetric complex Gaussian elements\cite{10243545,11018493}, i.e., $\boldsymbol{\Sigma}_{mk}[1,1]\sim\mathcal{CN}(0,\beta_{mk}\kappa/(\kappa+1))$
and $\boldsymbol{\Sigma}_{mk}[i,i]\sim\mathcal{CN}(0,\beta_{mk}/((\kappa+1)(L-1))),~i=2,3,...,L$\cite{10243545,11018493}. Here, $\kappa$ refers to the Ricean factor. $\beta_{mk}$ is the large-scale fading coefficient between the $m$-th AP and the $k$-th user, adopting the path loss model in \cite{7827017} with the relevant settings. The number of transmit and receive paths is the same, i.e., $L_{mk,r}=L_{mk,t}=L,~\forall m,k$\cite{11018493}. 
Until stated, FAS positions are located in a square region, where the length of the moving region side is $\lambda$, namely, $d_\text{max}-d_\text{min}=\lambda$ \cite{11018493}. $\kappa=1$ and $L=10$ are assumed for the response matrix.
The elevation and azimuth AoDs/AoAs are set to be i.i.d variables with the uniform distribution over $[-\pi/2,\pi/2]$ \cite{10243545}.
The carrier frequency is $f_c=2$ GHz, the communication bandwidth is $B=20$ MHz\cite{10103838}.
Meanwhile, the values of power consumption parameters are:
$p_{u}=20~\text{dBm}$, $\sigma^2=-91$ dBm, $\varepsilon_k=0.3,~\forall k$ \cite{10878991},  $P_{k,tc}=100$ mW, $P_{m,0}=100$ mW, $P_{m,AGC}=2$ mW, $P_{m,res}=10$ mW, $P_{m,bc}=0.25$ W/(Gbits/s), $f_s=2\times 10^9$ Hz, and $FOM_W=15$ fJ/conversion-step \cite{10858168,10878991,8097026}. For the AO-based algorithm, we set $b_\text{min}=1$, $b_\text{max}=5$ for bit allocation optimization, the maximum number of iterations $\text{I}_\text{max}=100$ and the tolerance $\epsilon=10^{-5}$.
\subsection{Effect of Quantization Bits}
\begin{figure}[t!]
    \centering
    \includegraphics[width=0.915\linewidth]{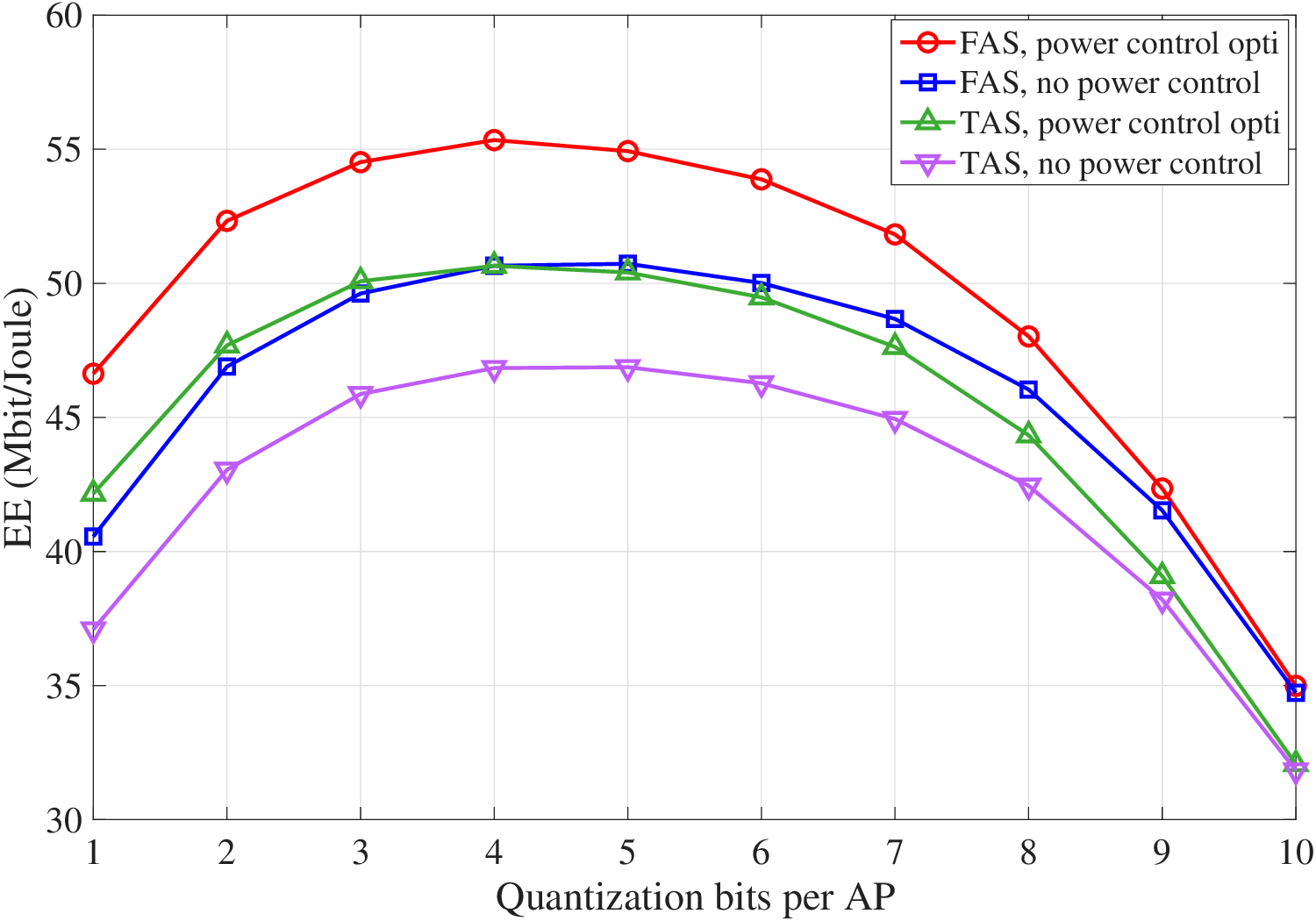}
    \caption{EE vs quantization bits per AP with $M=20$, $N=4$, $K=10$.}
    \label{Bit}
      \vspace{-10pt}
\end{figure}
\begin{figure}[t!]
    \centering
    \includegraphics[width=0.915\linewidth]{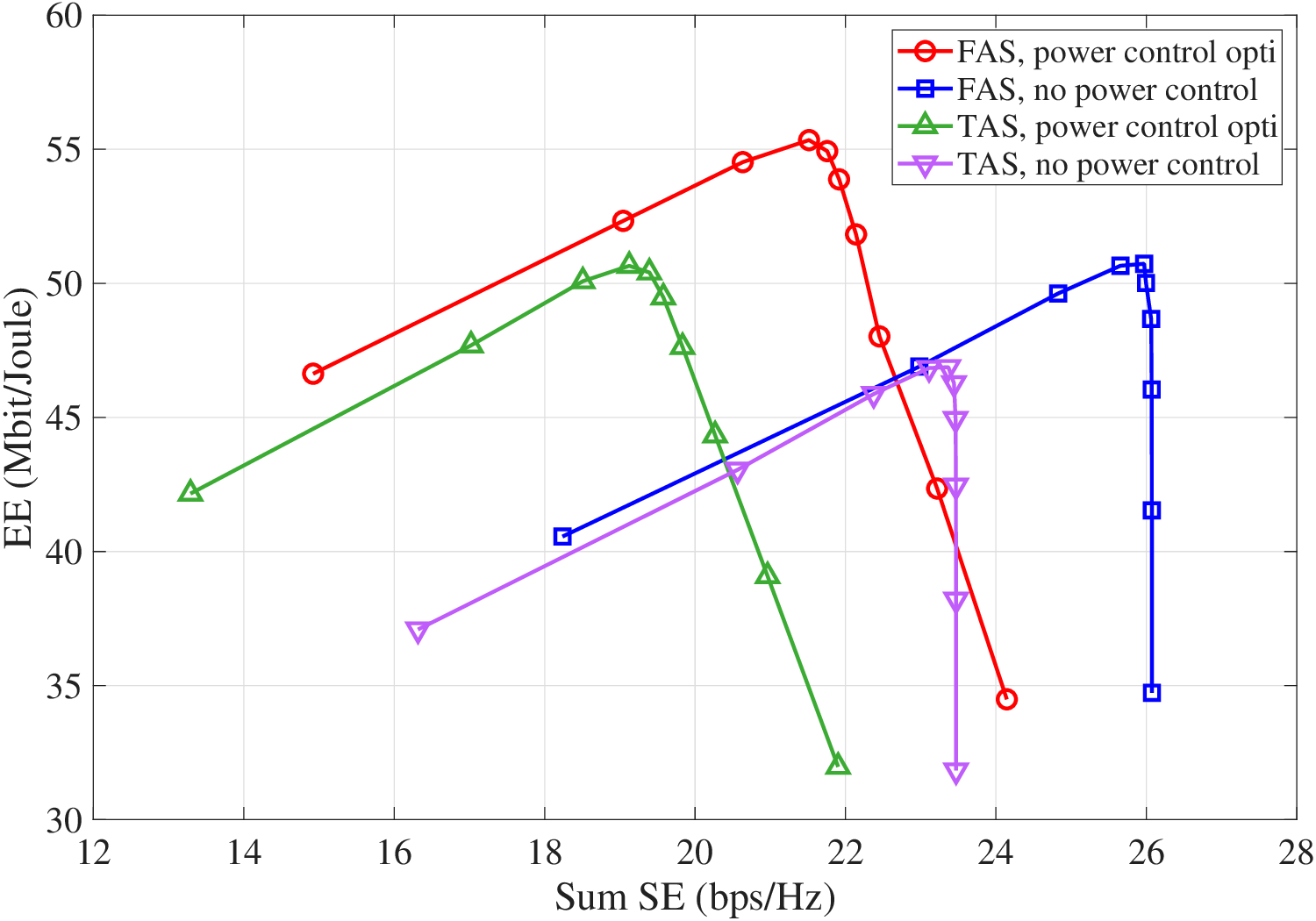}
    \caption{Trade-off between uplink sum SE and EE with $M=20$, $N=4$, $K=10$.}
    \label{Bit_SE}
    \vspace{-10pt}
\end{figure}
Fig. \ref{Bit} illustrates the EE performance versus the number of quantization bits per AP, where all APs share equal bit allocation, $b_m=b,~\forall m$. A prominent concave trend can be observed, indicating that an optimal EE among all feasible quantization bits exists. In the low-resolution regime, the EE increases monotonically since a higher quantization bit effectively mitigates signal distortion and improves signal strength, yielding substantial SE gains that far outweigh the marginal power consumption increase to improve EE. However, beyond the optimal threshold of 4 to 5 bits, as shown, the EE curves exhibit a sharp decline. This degradation occurs because the SE asymptotically approaches its unquantized saturation limit, whereas the power consumption with high resolutions in \eqref{P_mADC} scales up drastically, thereby reducing EE performance. Furthermore, the results clearly demonstrate that the application of FASs can introduce an additional $10\%$ EE increase over fixed-position antennas across the entire regime, validating that the spatial DoFs unlocked by FASs can reconfigure the propagation environment to enhance signal strength and improve system performance. Notably, the performance gain yielded by power control optimization exhibits a profound dependency on the ADC resolution. In the low and moderate-resolution regimes, the power control optimization achieves around $10\%$ EE increase over the no power control baseline, i.e., $\eta_k=1,~\forall k$. Conversely, in the high-resolution regime, the curves of optimized power control and no power control gradually merge because, as $b$ increases, the quantization noise becomes negligible, and the exponential growth of ADC power begins to play a dominant role. Consequently, the marginal increase in sum SE achieved by full-power transmission is insufficient to confront the massive hardware power, rendering the power control optimization ineffective from an EE perspective. Therefore, it is necessary to adopt FASs in cell-free massive MIMO systems with low-to-moderate resolution ADCs, coupled with efficient power control optimization, to attain the optimal EE configuration.

Fig. \ref{Bit_SE} illustrates the EE-SE trade-off trajectories parameterized by varying $b_m=b,~\forall m,$ from 1 bit to 10 bits, in 1-bit increments. In the low-resolution regime, increasing the ADC resolution yields a win-win scenario where both the sum SE and EE increase simultaneously, as the mitigation of quantization noise significantly boosts the sum SE faster than the growth in hardware power. However, a critical turning point occurs at $b=4\sim 5$, where the growth of the sum SE begins to slow down or hits a vertical saturation wall. Beyond this threshold, quantization error becomes negligible compared to multi-user interference, weakening the benefits of high resolutions in sum SE. Consequently, the exponential increase in high-resolution ADC power consumption dominates the total power denominator, causing a significant drop in EE.
Meanwhile, the scenarios with no power control achieve the absolute maximum sum SE but suffer severe EE penalties. In contrast, power control optimization acts as an EE-centric lever to boost EE performance, sacrificing a marginal fraction of sum SE. Furthermore, FASs consistently outperform fixed-position antennas, indicating that their spatial flexibilities provide a consistent SINR enhancement, thereby improving both SE and EE. Ultimately, the results validate the potential of 4-bit or 5-bit ADCs, coupled with the joint FAS position and power control optimization, as energy-efficient configurations without compromising the sum SE performance. 

\subsection{Effect of FAS Configuration}
\begin{figure}[t!]
    \centering
    \includegraphics[width=0.915\linewidth]{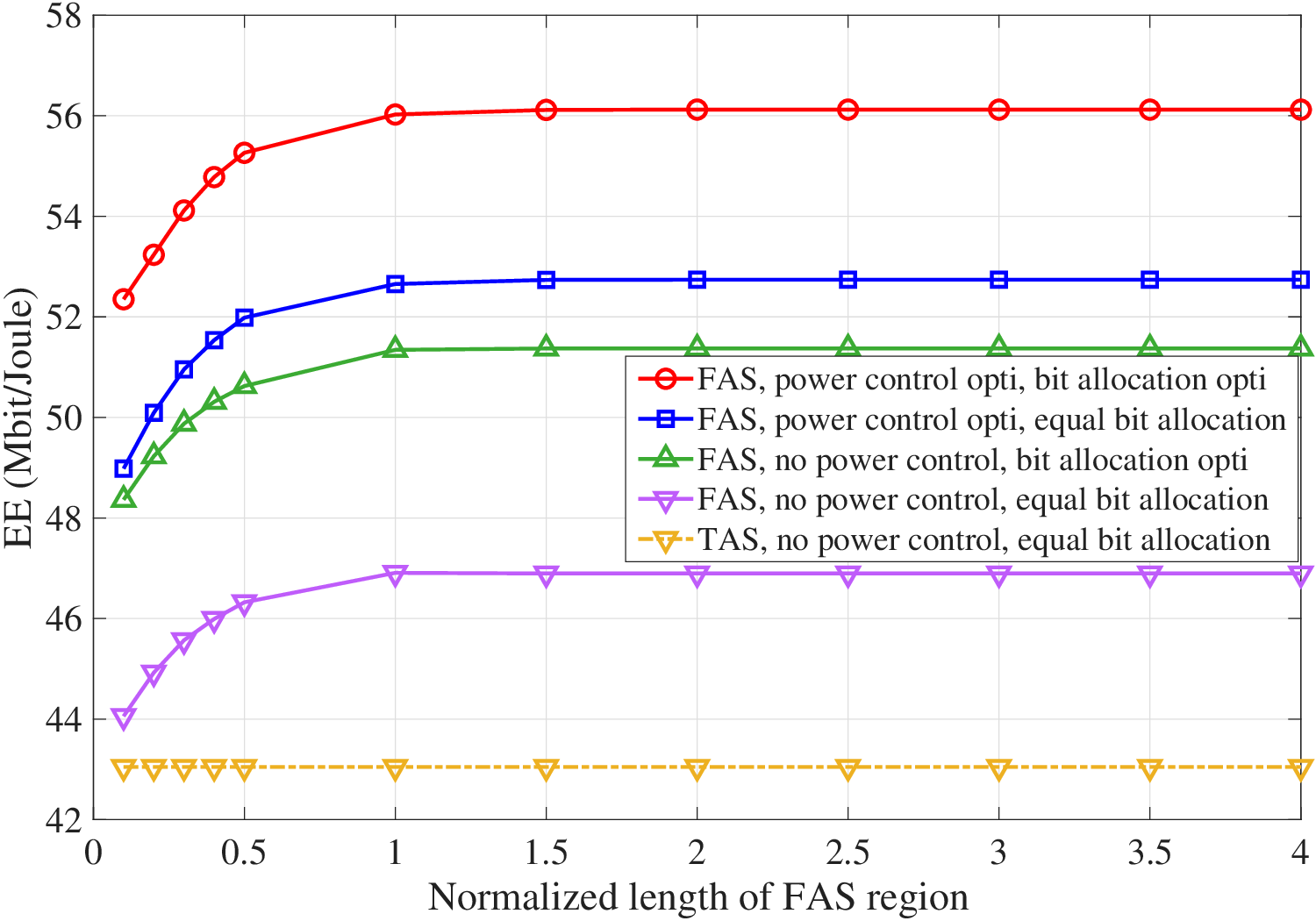}
    \caption{EE vs the normalized length of the FAS region side, $(d_\text{max}-d_\text{min})/\lambda$, with $M=20$, $N=4$, $K=10$.}
    \label{FAS}
      \vspace{-10pt}
\end{figure}
Fig. \eqref{FAS} illustrates EE as a function of the designated length of the FAS moving region side, $\left(d_\text{max}-d_\text{min}\right)/\lambda$. 
The results show that the EE exhibits a rapid initial ascent, followed by saturation as the moving region expands. For example, compared to $d_\text{max}-d_\text{min}=0.1\lambda$, $d_\text{max}-d_\text{min}=\lambda$ provides nearly $8\%$ EE increase. Meanwhile, the performance gains saturate for $d_\text{max}-d_\text{min}>\lambda$.
The rapid growth demonstrates that the FAS can exploit spatial diversity to drastically enhance system performance even with a slight mechanical adjustment. However, the saturation of EE performance indicates that expanding the moving region does not unlock additional spatial DoFs, suggesting that a moderately sized FAS moving region is sufficient to capture most of the available spatial DoFs in the proposed propagation environment. Meanwhile, FASs with $d_\text{max}-d_\text{min}=\lambda$ can introduce $10\%$ EE increase over fixed-position antennas, suggesting that FASs with optimized positions can utilize the additional spatial DoFs to natively advance the EE. Consequently, adopting highly compact, moderately sized FASs in cell-free massive MIMO systems mitigates the performance loss due to low-resolution ADCs, making their combination an appealing solution for boosting both EE and sum SE in 6G networks.
More specifically, referring to Sec. \ref{complexity}, the results reveal a comprehensive evaluation of the proposed framework by the iteration-performance trade-offs of each optimization subproblem. The Dinkelbach FP-based power control typically converges within merely 60 to 100 iterations; however, it benefits from low per-iteration complexity while yielding a 10\% marginal EE gain by aggressively muting redundant transmit power and managing inter-user interference. In contrast, the APGA-based FAS position optimization contributing 10\% marginal EE gain at $d_\text{max}-d_\text{min}=\lambda$ generally requires 10 to 30 iterations but incurs a higher per-iteration complexity due to gradient evaluation across multi-path components. Meanwhile, the APGA-based bit allocation imposes a moderate computational burden, requiring 20 to 40 iterations, and delivers roughly 10\% marginal EE gain by dynamically shifting resolutions toward dominant APs to offset the exponentially increasing ADC power consumption. Ultimately, the overall AO-based algorithm requires merely 5 to 10 outer iterations to coverage. While the subproblems have varying computational overheads, their synergistic combination delivers a 30\% EE increase. This reveals that despite the polynomial growth in complexity as the number of APs and users scales, the rapid outer-loop convergence justifies the computational cost. Meanwhile, the holistic joint optimization is crucial for maximizing the potential of FAS-enabled architectures with low-resolution ADCs. 

\subsection{Effect of Number of APs and Number of Users}
\begin{figure}[t!]
    \centering
    \includegraphics[width=0.915\linewidth]{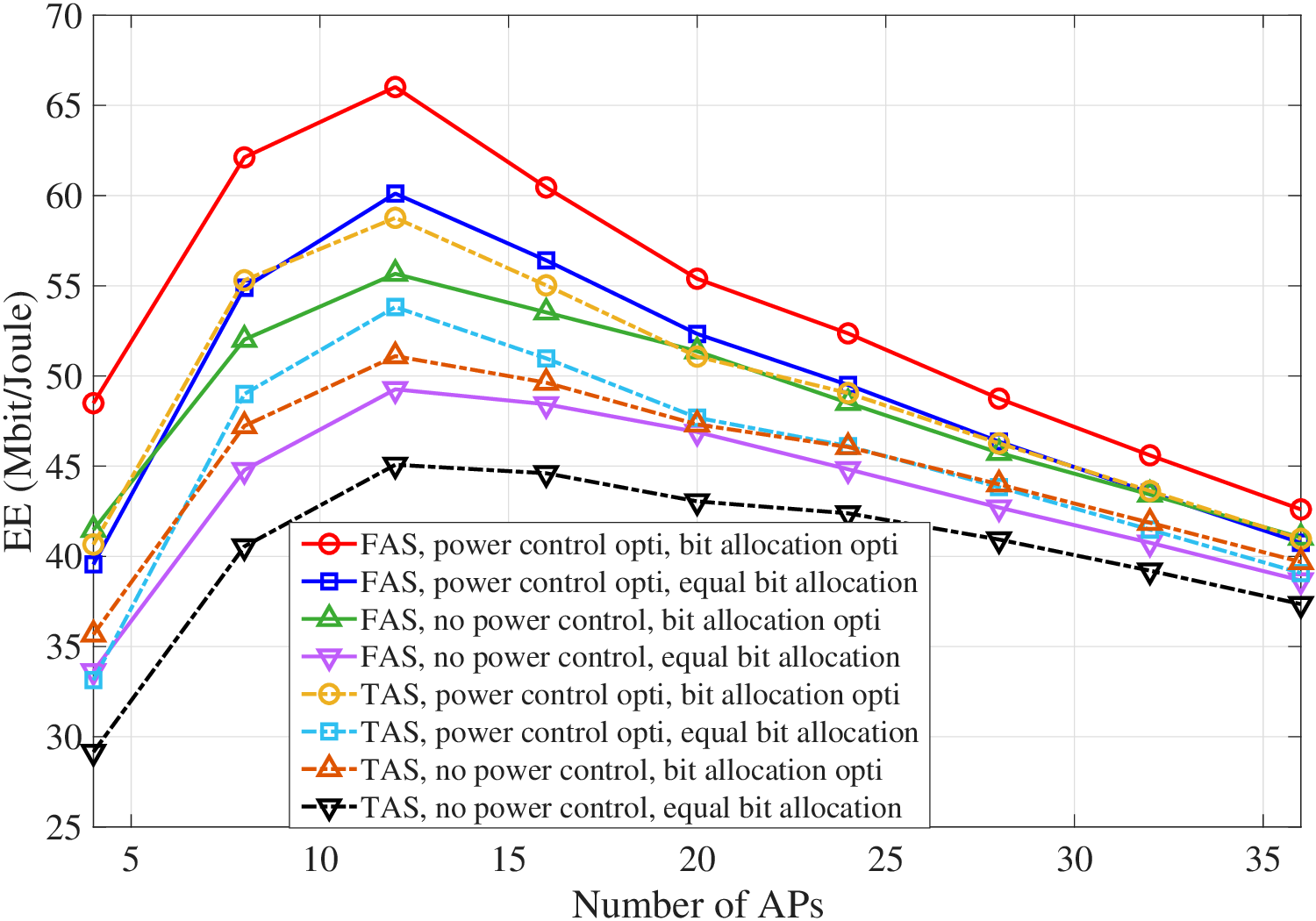}
    \caption{EE vs the number of APs with $N=4$, $K=10$.}
    \label{AP}
        \vspace{-2pt}
\end{figure}
\begin{figure}[t!]
    \centering
    \includegraphics[width=0.915\linewidth]{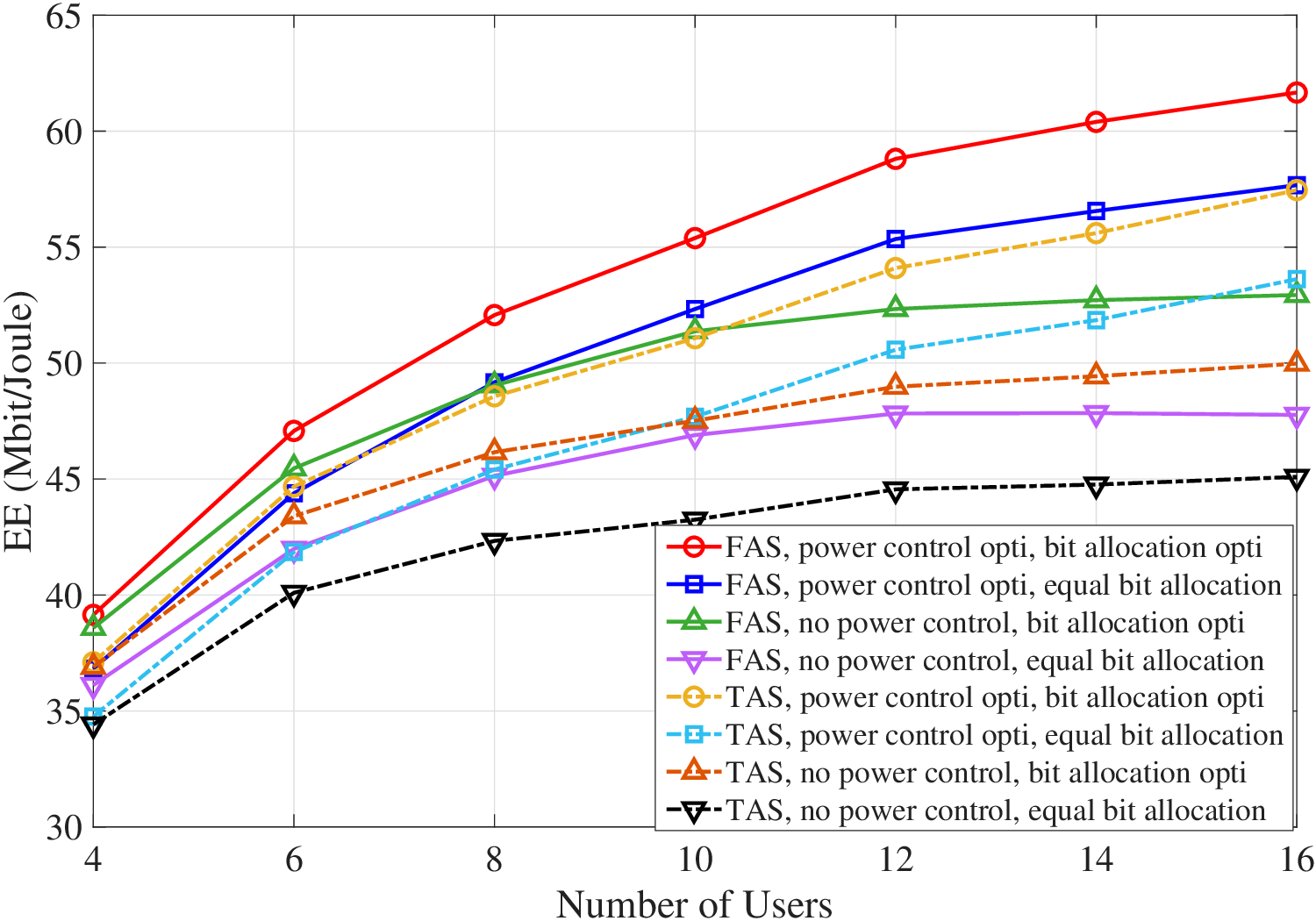}
    \caption{EE vs the number of users with $M=20$, $N=4$.}
    \label{user}
        \vspace{-12pt}
\end{figure}
Fig. \eqref{AP} shows that EE is a concave function of the number of APs, characterized by rapid initial ascent and a subsequent monotonic decline, indicating the fundamental EE trade-off inherent to distributed cell-free massive MIMO systems. 
When $M\leq 12$, introducing additional APs unlocks massive macro-diversity
gains, boosting the sum SE faster than hardware power growth, thereby improving EE. However, when $M\geq 12$, the marginal SE gain from adding new APs diminishes, while the circuit and ADC power consumption of new APs continues to scale linearly, dominating the total power and forcing the EE decrease. The results clearly demonstrate that the FAS position, power control and bit allocation optimizations can introduce a $3\sim15\%$, $4\sim15\%$ and $6\sim22\%$ EE increase over the unoptimized baseline, respectively. However, the growth rate introduced by these optimizations decreases as the number of APs increases, inducing a localized spatial hardening effect that gradually diminishes the benefits of the physical-layer and resource-layer optimizations. Fortunately, when joint optimization is adopted, a $15\sim66\%$ EE increase can be achieved, meeting the optimal EE and maintaining robust performance gains even with a larger number of APs. This indicates that FAS positions that can reconfigure the propagation environment, power control coefficients that can manage multi-user interference, and bit allocation that can reduce redundant quantization power at low-contribution APs should be optimized jointly to sustain energy-efficient communications in dense AP deployments.

Fig. \eqref{user} displays the relationship between EE and the number of users. The results show that EE increases monotonically with the number of users, driven by inherent multi-user spatial multiplexing gains. This is because serving more users significantly boosts the sum SE, effectively offsetting the massive power consumption of AP circuits and ADCs.
The results show that without power control, i.e., $\eta_k=1,~\forall k$, the EE gradually saturates as the number of users increases, since dense deployments introduce higher multi-user interference, which bottlenecks the growth of the sum SE. In contrast, the aggregate user transmit power continues to scale linearly, halting further EE gains. 
Conversely, the power control optimization continuously maintains an increasing EE trajectory, i.e., $1\sim18\%$ EE increase compared to the no power control baseline, by actively compressing user transmit power and preventing severe multi-user interference.
Furthermore, the application of FASs consistently outperforms fixed-position antennas by $5\sim10\%$ EE increase across the entire regime, leveraging increased spatial reconfigurability and spatial DoFs to improve system performance. Meanwhile, bit allocation optimization dynamically assigns higher resolutions to dominant APs to introduce around $9\%$ EE increase, balancing the trade-off between the exponential ADC power and the sum SE.
 Ultimately, the joint optimization can achieve $13\sim36\%$ EE increase, indicating that supporting massive connectivity in energy-efficient 6G networks requires a holistic optimization. Referring to Sec. \ref{complexity} and Fig. \eqref{AP}, as the number of APs and users grows, the joint optimization experiences
increasing computational complexity. To maintain scalability and
low-complexity computational feasibility in practice, our future work will integrate this FAS-enabled framework with efficient user-centric architectures, alongside other low-complexity signal processing
approaches.

\section{Conclusion}

This paper proposed a novel, energy-efficient cell-free massive MIMO system that synergistically integrates FASs with low-resolution ADCs. To rigorously characterize this system, we developed a comprehensive uplink transmission model. We also derived analytical SE and EE expressions under local MMSE processing, explicitly capturing the quantization errors based on AQNM. To maximize EE, we proposed an AO-based algorithm. This framework employs the Dinkelbach algorithm-based FP for power control, alongside novel APGA algorithms for optimizing FAS positions and ADC bit allocations.
Numerical results verified that low-resolution ADCs aggressively compress signals to save hardware power, inevitably degrading the sum SE while maintaining EE performance. However, FASs can recover this SE loss by steering the antenna to locations that can highly improve received signal
levels prior to destructive quantization. Consequently, FASs compensate for the performance loss of low-resolution ADCs without requiring additional transmit power. This makes their combination an appealing solution for boosting both EE and sum SE in 6G networks.
Furthermore, our analysis revealed that maximizing EE in dense networks necessitates efficient resource allocations. Power control acts as both an energy-saving mechanism and an interference-management lever. Concurrently, efficient bit allocation mitigates the exponential growth in ADC power by allocating higher resolutions to dominant APs. Ultimately, this work establishes that achieving energy-efficient 6G networks requires a holistic joint optimization, with FAS to unlock additional spatial DoFs, power control to suppress interference, and bit allocation to manage hardware power. Moreover, a low-complexity approach to realizing the aforementioned benefits in ultra-dense 6G networks will be a nontrivial topic in our future work.

\ifCLASSOPTIONcaptionsoff
 \newpage
\fi


%




\ifCLASSOPTIONcaptionsoff
 \newpage
\fi



\bibliographystyle{IEEEtran}
\bibliography{IEEEabrv,ref}

\begin{thebibliography}{10}
\providecommand{\url}[1]{#1}
\csname url@samestyle\endcsname
\providecommand{\newblock}{\relax}
\providecommand{\bibinfo}[2]{#2}
\providecommand{\BIBentrySTDinterwordspacing}{\spaceskip=0pt\relax}
\providecommand{\BIBentryALTinterwordstretchfactor}{4}
\providecommand{\BIBentryALTinterwordspacing}{\spaceskip=\fontdimen2\font plus
\BIBentryALTinterwordstretchfactor\fontdimen3\font minus \fontdimen4\font\relax}
\providecommand{\BIBforeignlanguage}[2]{{%
\expandafter\ifx\csname l@#1\endcsname\relax
\typeout{** WARNING: IEEEtran.bst: No hyphenation pattern has been}%
\typeout{** loaded for the language `#1'. Using the pattern for}%
\typeout{** the default language instead.}%
\else
\language=\csname l@#1\endcsname
\fi
#2}}
\providecommand{\BIBdecl}{\relax}
\BIBdecl

\bibitem{7827017}
H.~Q. Ngo, A.~Ashikhmin, H.~Yang, E.~G. Larsson, and T.~L. Marzetta, ``Cell-free massive {MIMO} versus small cells,'' \emph{IEEE Trans. Wireless Commun.}, vol.~16, no.~3, pp. 1834--1850, 2017.

\bibitem{9665300}
T.~Van~Chien, H.~Q. Ngo, S.~Chatzinotas, M.~Di~Renzo, and B.~Ottersten, ``Reconfigurable intelligent surface-assisted cell-free massive {MIMO} systems over spatially-correlated channels,'' \emph{IEEE Trans. Wireless Commun.}, vol.~21, no.~7, pp. 5106--5128, 2022.

\bibitem{8097026}
H.~Q. Ngo, L.-N. Tran, T.~Q. Duong, M.~Matthaiou, and E.~G. Larsson, ``On the total energy efficiency of cell-free massive {MIMO},'' \emph{IEEE Trans. Green Commun. Netw.}, vol.~2, no.~1, pp. 25--39, 2018.

\bibitem{7917284}
E.~Nayebi, A.~Ashikhmin, T.~L. Marzetta, H.~Yang, and B.~D. Rao, ``Precoding and power optimization in cell-free massive {MIMO} systems,'' \emph{IEEE Trans. Wireless Commun.}, vol.~16, no.~7, pp. 4445--4459, 2017.

\bibitem{11196010}
J.~Qian, R.~Murch, and K.~B. Letaief, ``Performance analysis of {STAR-RIS}-assisted cell-free massive {MIMO} systems with electromagnetic interference and phase errors,'' \emph{IEEE Trans. Wireless Commun.}, vol.~25, pp. 4982--5000, 2026.

\bibitem{10032129}
J.~Zheng, J.~Zhang, J.~Cheng, V.~C.~M. Leung, D.~W.~K. Ng, and B.~Ai, ``Asynchronous cell-free massive {MIMO} with rate-splitting,'' \emph{IEEE J. Sel. Areas Commun.}, vol.~41, no.~5, pp. 1366--1382, 2023.

\bibitem{9875036}
A.~Papazafeiropoulos, I.~Krikidis, and P.~Kourtessis, ``Impact of channel aging on reconfigurable intelligent surface aided massive {MIMO} systems with statistical {CSI},'' \emph{IEEE Trans. Veh. Technol.}, vol.~72, no.~1, pp. 689--703, 2023.

\bibitem{8845768}
E.~Björnson and L.~Sanguinetti, ``Making cell-free massive {MIMO} competitive with {MMSE} processing and centralized implementation,'' \emph{IEEE Trans. Wireless Commun.}, vol.~19, no.~1, pp. 77--90, 2020.

\bibitem{10858168}
Y.~Zhang, W.~Xia, H.~Zhao, Y.~Mao, J.~Zhang, and G.~Zheng, ``Enhancing uplink performance for cell-free massive {MIMO} with low-resolution {ADCs} by {RSMA},'' \emph{IEEE Journal on Selected Areas in Communications}, vol.~43, no.~3, pp. 720--735, 2025.

\bibitem{10878991}
Y.~Xiong, L.~Tang, S.~Sun, S.~Yang, L.~Liu, S.~Mao, Z.~Zhang, and N.~Wei, ``{RIS}-aided cell-free massive {MIMO} systems with low-resolution {ADCs}: Uplink performance analysis and optimization,'' \emph{IEEE Internet of Things Journal}, pp. 1--1, 2025.

\bibitem{8756265}
X.~Hu, C.~Zhong, X.~Chen, W.~Xu, H.~Lin, and Z.~Zhang, ``Cell-free massive {MIMO} systems with low resolution {ADCs},'' \emph{IEEE Trans. Commun.}, vol.~67, no.~10, pp. 6844--6857, 2019.

\bibitem{8811486}
Y.~Zhang, M.~Zhou, X.~Qiao, H.~Cao, and L.~Yang, ``On the performance of cell-free massive {MIMO} with low-resolution {ADCs},'' \emph{IEEE Access}, vol.~7, pp. 117\,968--117\,977, 2019.

\bibitem{9799777}
I.-s. Kim, M.~Bennis, and J.~Choi, ``Cell-free mmwave massive {MIMO} systems with low-capacity fronthaul links and low-resolution {ADC/DACs},'' \emph{IEEE Transactions on Vehicular Technology}, vol.~71, no.~10, pp. 10\,512--10\,526, 2022.

\bibitem{9318477}
Y.~Zhang, L.~Yang, and H.~Zhu, ``Cell-free massive {MIMO} systems with low-resolution {ADCs}: The rician fading case,'' \emph{IEEE Systems Journal}, vol.~16, no.~1, pp. 1471--1482, 2022.

\bibitem{10041946}
X.~Ma, X.~Lei, P.~T. Mathiopoulos, K.~Yu, and X.~Tang, ``Scalable cell-free massive {MIMO} systems with finite resolution {ADCs/DACs} over spatially correlated rician fading channels,'' \emph{IEEE Transactions on Vehicular Technology}, vol.~72, no.~6, pp. 7699--7716, 2023.

\bibitem{10902611}
P.~Liu, J.~Cui, Z.~Leng, J.~Hu, D.~Kong, K.~Wang, G.~Interdonato, and S.~Buzzi, ``Power allocation for cell-free massive {MIMO} two-way relay systems with low-resolution {ADCs},'' \emph{IEEE Trans. Commun.}, pp. 1--1, 2025.

\bibitem{9650760}
K.-K. Wong and K.-F. Tong, ``Fluid antenna multiple access,'' \emph{IEEE Trans. Wireless Commun.}, vol.~21, no.~7, pp. 4801--4815, 2022.

\bibitem{10318134}
W.~K. New, K.-K. Wong, H.~Xu, K.-F. Tong, C.-B. Chae, and Y.~Zhang, ``Fluid antenna system enhancing orthogonal and non-orthogonal multiple access,'' \emph{IEEE Commun. Lett.}, vol.~28, no.~1, pp. 218--222, 2024.

\bibitem{10146262}
K.-K. Wong, K.-F. Tong, and C.-B. Chae, ``Fluid antenna system—part {III}: A new paradigm of distributed artificial scattering surfaces for massive connectivity,'' \emph{IEEE Commun. Lett.}, vol.~27, no.~8, pp. 1929--1933, 2023.

\bibitem{10328751}
Y.~Ye, L.~You, J.~Wang, H.~Xu, K.-K. Wong, and X.~Gao, ``Fluid antenna-assisted {MIMO} transmission exploiting statistical {CSI},'' \emph{IEEE Commun. Lett.}, vol.~28, no.~1, pp. 223--227, 2024.

\bibitem{10740058}
J.~Zhang, J.~Rao, Z.~Li, Z.~Ming, C.-Y. Chiu, K.-K. Wong, K.-F. Tong, and R.~Murch, ``A novel pixel-based reconfigurable antenna applied in fluid antenna systems with high switching speed,'' \emph{IEEE Open J. Antennas Propag.}, vol.~6, no.~1, pp. 212--228, 2025.

\bibitem{10066316}
K.-K. Wong, D.~Morales-Jimenez, K.-F. Tong, and C.-B. Chae, ``Slow fluid antenna multiple access,'' \emph{IEEE Trans. Commun.}, vol.~71, no.~5, pp. 2831--2846, 2023.

\bibitem{10103838}
M.~Khammassi, A.~Kammoun, and M.-S. Alouini, ``A new analytical approximation of the fluid antenna system channel,'' \emph{IEEE Trans. Wireless Commun.}, vol.~22, no.~12, pp. 8843--8858, 2023.

\bibitem{10707252}
L.~Zhou, J.~Yao, M.~Jin, T.~Wu, and K.-K. Wong, ``Fluid antenna-assisted {ISAC} systems,'' \emph{IEEE Wireless Commun. Lett.}, vol.~13, no.~12, pp. 3533--3537, 2024.

\bibitem{10694457}
M.~Olyaee and S.~Buzzi, ``User-centric cell-free massive {MIMO} with access points empowered by fluid antennas,'' in \emph{Proc. IEEE SPAWC}, 2024, pp. 666--670.

\bibitem{10967080}
T.~Han, Y.~Zhu, K.-K. Wong, G.~Zheng, and H.~Shin, ``Cell-free fluid antenna multiple access networks,'' \emph{IEEE Trans. Wireless Commun.}, pp. 1--1, 2025.

\bibitem{10891142}
X.~Shi, X.~Shao, B.~Zheng, and R.~Zhang, ``{6DMA}-aided cell-free massive {MIMO} communication,'' \emph{IEEE Wireless Communications Letters}, vol.~14, no.~5, pp. 1361--1365, 2025.

\bibitem{10827177}
J.~Guan, B.~Lyu, Y.~Liu, and F.~Tian, ``Secure transmission for movable antennas empowered cell-free symbiotic radio communications,'' in \emph{Proc. IEEE WCSP}, 2024, pp. 578--584.

\bibitem{11018493}
H.~Wei, W.~Wang, W.~Ni, C.~Zhang, and Y.~Huang, ``Movable-antenna enabled cell-free networks,'' \emph{IEEE Transactions on Vehicular Technology}, pp. 1--6, 2025.

\bibitem{NIPS2015_f7664060}
\BIBentryALTinterwordspacing
H.~Li and Z.~Lin, ``Accelerated proximal gradient methods for nonconvex programming,'' in \emph{Advances in Neural Information Processing Systems}, C.~Cortes, N.~Lawrence, D.~Lee, M.~Sugiyama, and R.~Garnett, Eds., vol.~28.\hskip 1em plus 0.5em minus 0.4em\relax Curran Associates, Inc., 2015. [Online]. Available: \url{https://proceedings.neurips.cc/paper_files/paper/2015/file/f7664060cc52bc6f3d620bcedc94a4b6-Paper.pdf}
\BIBentrySTDinterwordspacing

\bibitem{9217298}
M.~Farooq, H.~Q. Ngo, and L.~N. Tran, ``Accelerated projected gradient method for the optimization of cell-free massive {MIMO} downlink,'' in \emph{Proc. IEEE 2020 PIMRC}, 2020, pp. 1--6.

\bibitem{8388873}
J.~Qian, C.~Masouros, and A.~Garcia-Rodriguez, ``Partial {CSI} acquisition for size-constrained massive {MIMO} systems with user mobility,'' \emph{IEEE Trans. Veh. Technol.}, vol.~67, no.~9, pp. 9016--9020, 2018.

\bibitem{10318061}
L.~Zhu, W.~Ma, and R.~Zhang, ``Modeling and performance analysis for movable antenna enabled wireless communications,'' \emph{IEEE Trans. Wireless Commun.}, vol.~23, no.~6, pp. 6234--6250, 2024.

\bibitem{11016053}
T.~Hao, C.~Shi, Q.~Wu, B.~Xia, Y.~Guo, L.~Ding, and F.~Yang, ``Fluid-antenna enhanced isac: Joint antenna positioning and dual-functional beamforming design under perfect and imperfect csi,'' \emph{IEEE Transactions on Vehicular Technology}, pp. 1--16, 2025.

\bibitem{10354003}
L.~Zhu, W.~Ma, B.~Ning, and R.~Zhang, ``Movable-antenna enhanced multiuser communication via antenna position optimization,'' \emph{IEEE Transactions on Wireless Communications}, vol.~23, no.~7, pp. 7214--7229, 2024.

\bibitem{10243545}
W.~Ma, L.~Zhu, and R.~Zhang, ``Mimo capacity characterization for movable antenna systems,'' \emph{IEEE Transactions on Wireless Communications}, vol.~23, no.~4, pp. 3392--3407, 2024.

\bibitem{9123382}
G.~Femenias and F.~Riera-Palou, ``Fronthaul-constrained cell-free massive {MIMO} with low resolution {ADCs},'' \emph{IEEE Access}, vol.~8, pp. 116\,195--116\,215, 2020.

\bibitem{10755170}
B.~Xu, Y.~Chen, Q.~Cui, X.~Tao, and K.-K. Wong, ``Sparse bayesian learning-based channel estimation for fluid antenna systems,'' \emph{IEEE Wireless Communications Letters}, vol.~14, no.~2, pp. 325--329, 2025.

\bibitem{10236898}
W.~Ma, L.~Zhu, and R.~Zhang, ``Compressed sensing based channel estimation for movable antenna communications,'' \emph{IEEE Communications Letters}, vol.~27, no.~10, pp. 2747--2751, 2023.

\bibitem{11091342}
Y.~Xiong, L.~Tang, S.~Yang, S.~Sun, and L.~Liu, ``Cell-free massive {MIMO} networks with low-resolution quantization: Downlink performance analysis and optimization,'' \emph{IEEE J. Sel. Top. Signal Process.}, vol.~19, no.~6, pp. 1055--1071, 2025.

\bibitem{9205899}
A.~Kaushik, E.~Vlachos, C.~Tsinos, J.~Thompson, and S.~Chatzinotas, ``Joint bit allocation and hybrid beamforming optimization for energy efficient millimeter wave {MIMO} systems,'' \emph{IEEE Trans. Green Commun. Netw.}, vol.~5, no.~1, pp. 119--132, 2021.

\bibitem{8314727}
K.~Shen and W.~Yu, ``Fractional programming for communication systems—part {I}: Power control and beamforming,'' \emph{IEEE Trans. Signal Proc.}, vol.~66, no.~10, pp. 2616--2630, 2018.

\bibitem{9519163}
H.~A. Ammar, R.~Adve, S.~Shahbazpanahi, G.~Boudreau, and K.~V. Srinivas, ``Downlink resource allocation in multiuser cell-free {MIMO} networks with user-centric clustering,'' \emph{IEEE Trans. Wireless Commun.}, vol.~21, no.~3, pp. 1482--1497, 2022.

\bibitem{10256066}
M.~Sarker and A.~O. Fapojuwo, ``Fractional programming-based uplink transmission power allocation for user-centric cell-free massive {MIMO} systems,'' \emph{IEEE Trans. Green Commun.}, vol.~8, no.~1, pp. 50--63, 2024.

\bibitem{9293031}
S.~Chakraborty, O.~T. Demir, E.~Bj\"{o}rnson, and P.~Giselsson, ``Efficient downlink power allocation algorithms for cell-free massive {MIMO} systems,'' \emph{OJ-COMS}, vol.~2, pp. 168--186, 2021.

\bibitem{10812717}
H.~Fang, H.~Hu, B.~Wu, Y.~Zhang, L.~Yang, and H.~Zhu, ``Power allocation and precoding design for active {RIS}-aided cell-free massive {MIMO} systems,'' \emph{IEEE Internet of Things Journal}, pp. 1--1, 2024.

\bibitem{9709200}
T.~C. Mai, H.~Q. Ngo, and L.-N. Tran, ``Energy efficiency maximization in large-scale cell-free massive {MIMO}: A projected gradient approach,'' \emph{IEEE Trans. Wireless Commun.}, vol.~21, no.~8, pp. 6357--6371, 2022.

\bibitem{hu2025uplinktransmissiondesignfluid}
\BIBentryALTinterwordspacing
L.~Hu, L.~Li, C.~Pan, and H.~Ren, ``Uplink transmission design for fluid antenna-enabled multiuser {MIMO} systems with imperfect {CSI},'' 2025. [Online]. Available: \url{https://arxiv.org/abs/2503.01668}
\BIBentrySTDinterwordspacing

\bibitem{10388242}
G.~Hu, Q.~Wu, K.~Xu, J.~Ouyang, J.~Si, Y.~Cai, and N.~Al-Dhahir, ``Fluid antennas-enabled multiuser uplink: A low-complexity gradient descent for total transmit power minimization,'' \emph{IEEE Communications Letters}, vol.~28, no.~3, pp. 602--606, 2024.

\end{thebibliography}
\end{document}